\documentclass[11pt,a4paper]{article}
\usepackage{jheppub}

\usepackage{mathrsfs}
\usepackage{url}
 
\newcommand{\real}{\ensuremath{\mathbb{R}}}

\newcommand{\tr}{\ensuremath{\text{tr}}}

\newcommand{\bigo}{\ensuremath{\mathcal{O}}}
\newcommand{\smallo}{\ensuremath{o}}

\newcommand{\barr}[1]{\ensuremath{\overline{#1}}}

\newcommand{\ad}{\ensuremath{\text{ad}\,}}

\newcommand{\eqdef}{\ensuremath{:=}}
\newcommand{\defeq}{\ensuremath{=:}}
\newcommand{\su}{\ensuremath{\mathfrak{su}}}
\newcommand{\sprod}[2]{\ensuremath{{#1} \cdot {#2}}}
\newcommand{\extprod}[2]{\ensuremath{{#1} \times {#2}}}
\newcommand{\extder}{\ensuremath{\mathbf{d}}}
\newcommand{\weq}{\ensuremath{\approx}}
\newcommand{\liephase}{\ensuremath{\boldsymbol{\mathcal{L}}}}
\newcommand{\Sym}{\mathsf{Sym}}
\newcommand{\Gau}{\mathsf{Gau}}
\newcommand{\Asym}{\mathsf{Asym}}
\newcommand{\SU}{\mathrm{SU}}
\newcommand{\eff}{\mathcal{F}}
\begin{document}
% \title{Asymptotic symmetries of Yang-Mills in the Hamiltonian formulation}
% \title{Study of the asymptotic symmetries of free Yang-Mills with the Hamiltonian formalism}
\title{Asymptotic symmetries of Yang-Mills fields\\ 
in Hamiltonian formulation}
\author[a]{Roberto Tanzi}%\footnote{Corresponding author: v}}
\author[a,b]{and Domenico Giulini}
\affiliation[a]{University of Bremen, Center of Applied Space Technology and Microgravity (ZARM), 28359 Bremen}
\affiliation[b]{Leibniz University of Hannover, Institute for Theoretical Physics, 30167 Hannover, Germany}
\emailAdd{roberto.tanzi@zarm.uni-bremen.de}
\emailAdd{giulini@itp.uni-hannover.de}
\arxivnumber{2006.07268}

\dedicated{Dedicated to the memory of Federico Tonielli.}

\abstract{
 We investigate the asymptotic symmetry group 
of the free $\SU(N)$-Yang-Mills theory using the 
Hamiltonian formalism. We closely follow the 
strategy  of Henneaux and Troessaert who successfully 
applied the Hamiltonian formalism to the case of 
gravity and electrodynamics, thereby deriving the 
respective asymptotic symmetry groups of these 
theories from clear-cut first principles. 
These principles include the minimal assumptions 
that are necessary to ensure the existence of 
Hamiltonian structures (phase space, symplectic
form, differentiable Hamiltonian) and, in 
case of Poincar\'e invariant theories, 
a canonical action of the  Poincar\'e group.
In the first part of the paper we show how these 
requirements can be met in the non-abelian 
$\SU(N)$-Yang-Mills case by imposing suitable 
fall-off and parity conditions on the fields.
We observe that these conditions admit neither
non-trivial asymptotic symmetries nor non-zero 
global charges. In the second part of the paper 
we discuss possible gradual relaxations of these 
conditions by following the same strategy that 
Henneaux and Troessaert had employed to remedy 
a similar situation in the electromagnetic case.
Contrary to our expectation and the findings of 
Henneaux and Troessaert for the abelian case, 
there seems to be no relaxation that meets 
the requirements of a Hamiltonian formalism 
\emph{and} allows for non-trivial asymptotic 
symmetries and charges. Non-trivial asymptotic
symmetries and charges are only possible if 
either the Poincar\'e group fails to act 
canonically or if the formal expression for the 
symplectic form diverges, i.e. the form 
does not exist.  This seems to hint at 
a kind of colour-confinement built into 
the classical Hamiltonian formulation of
non-abelian gauge theories.  
}

\maketitle

% \newpage
% \tableofcontents
% \newpage

\section{Introduction} \label{sec:introduction}

Asymptotic symmetries are those symmetries that appear 
in theories with long-ranging fields, such as 
gravity and electrodynamics. They appear in the 
formalism once the analytic behaviour of fields near 
infinity is specified. 
Although the first studies concerning general relativity 
at null infinity appeared more than half a century
ago~\cite{BMS1,BMS2,BMS3}, the subject have been 
revitalised more recently after it has been conjectured
that it may be related to the solution of the 
long-standing information-loss paradox~\cite{HPS} and 
it has been a very active area of research in the last 
years. Several studies have already analysed many aspects
of the topic, such as the situation at null infinity 
and the connection to soft
theorems~\cite{Strominger1,Strominger2,Strominger3,Strominger4,Strominger5,Strominger6,Campiglia1,Campiglia2,Conde,Heissenberg},  
the relation with the potentially-detectable memory effect~\cite{Compere1,Compere2,Francia}, the asymptotic (A)dS case~\cite{BMS-ADS1,BMS-ADS2}, and the situation 
at spatial infinity~\cite{Henneaux-GR,Henneaux-ED,Henneaux-ED-higher,Henneaux-ED-GR,Henneaux-scalar,Henneaux-review}.
It is in particular the latter that deeply connects with the present paper.

The study of asymptotic symmetries at spatial infinity 
uses the machinery of the Hamiltonian formulation of classical field theories and is complementary to the analogous studies at null infinity, which appeared chronologically sooner and are, perhaps, less demanding 
on the computational side.
The reason why one wishes, nevertheless, to study also the Hamiltonian treatment of the problem is not only that one should find the equivalence of the two approaches, but more importantly, that the Hamiltonian tools are very well suited for a systematic characterisation of state 
spaces and the symmetries it supports.
Needless to emphasise that it also  provides the 
basis for the canonical quantisation of the theory.

The systematic Hamiltonian study of asymptotic symmetries
was started by Henneaux and Troessaert, whose analysis covered a plethora of aspects: they analysed the case of general relativity~\cite{Henneaux-GR}, electrodynamics in four~\cite{Henneaux-ED} and higher dimensions~\cite{Henneaux-ED-higher}, the coupled Maxwell-Einstein theory~\cite{Henneaux-ED-GR}, and the massless scalar field~\cite{Henneaux-scalar}. 
The purpose of the present paper is to include 
non-abelian gauge fields in this list, which means 
to study special-relativistic $\SU(N)$-Yang-Mills 
theory in a proper Hamiltonian setting. This 
requires, first of all, the following basic 
structures to exist:
\begin{itemize}
\itemsep=-4pt
\item[(i)] 
a phase space;
\item[(ii)] 
a symplectic form on phase space; 
\item[(iii)]
a Hamiltonian as a differentiable 
function on phase space;
\item[(iv)]
a symplectic (or even Hamiltonian) action of 
the Poincar\'e group on phase space.
\end{itemize}
Regarding the last point, we recall that the 
action is symplectic or ``canonical'', if it 
preserves the symplectic structure. It is 
Hamiltonian if, in addition, Poincar\'e 
transformations on phase space are 
generated by phase-space functions,  giving rise
to globally defined Hamiltonian vector fields,
whose Poisson brackets form a faithful 
representation of the Lie algebra of the 
Poincar\'e group. This is also known as a 
\emph{comoment} for the action of the group;
compare, e.g.,  \cite[Chap.\,3.2]{Woodhouse:GQ}.
For general Lie groups there may be obstructions 
to turn a symplectic action into a Hamiltonian 
action (i.e. against the existence of a comoment), 
and even if the latter exists, it need not be 
unique. These issues of existence and uniqueness
are classified by the Lie algebra's second and 
first cohomology group, respectively. In case of 
the Poincar\'e group, these cohomology groups 
are both trivial, and these issues do not 
arise; compare,  e.g., \cite[Chap.\,3.3]{Woodhouse:GQ}.
In that case it is sufficient to demand a symplectic
or, as we will henceforth say, canonical action. 
 
It should be clear that the possibility to simultaneously meet the requirements (i-iv) listed above will delicately depend on the precise characterisation of phase space.
For field theories this entails to characterise the canonical fields in terms of fall-off conditions and, as it turns out, also parity conditions.
The former ones tell us how quickly the fields vanish as one approaches spatial infinity, whereas the latter ones tell us the parity of the leading term in the asymptotic expansion of the fields as functions on the 2-sphere at spatial infinity. 
In the context of Hamiltonian general relativity it has long been realised that parity conditions are necessary in order to ensure the existence of integrals that represent Hamiltonian generators of symmetries that one wishes to include on field 
configurations that are asymptotically Minkowskian and represent isolated systems; compare \cite{RT,Beig}.

Quite generally, the task is to find a compromise 
between two competing aspects: the size of phase 
space and the implementation of symmetries. On the
one hand, phase space should be large enough to 
contain sufficiently many interesting states, in 
particular those being represented by fields whose
asymptotic fall-off is slow enough to allow globally 
`charged' states, like electric charge for the 
Coulomb solution in Electrodynamics,  or mass for 
the Schwarzschild solution in General Relativity. 
On the other hand, for the  symmetry generators 
to exist as (differentiable) Hamiltonian functions,
phase space cannot be too extensive. 
Since we are dealing with relativistic theories,
the compatible symmetries should contain the 
Poincar\'e group, but might likely turn out to be 
a non-trivial extension thereof if we are dealing 
with gauge or diffeomorphism-invariant theories.

Let us illustrate this last point in a somewhat more
mathematical language. In any gauge- or 
diffeomorphism-invariant theory, there is a large,
infinite-dimensional group acting on the fields which
transforms solutions of the equations of motions 
to solutions (of the very same equations). 
For example, in ordinary gauge theories, these 
are certain (infinite-dimensional) groups of 
bundle automorphisms, or, in general relativity, 
the group of diffeomorphisms of some smooth manifold. 
Let us call it the ``symmetry group'' $\Sym$. Now, inside 
$\Sym$, there is a normal subgroup of ``gauge transformations'',
denoted by $\Gau$. They, too, are symmetries in the 
sense that they map solutions of the field equations 
to solutions, but they are distinguished by their
interpretation as ``redundancies in description''.
This means that any two phase-space points 
connected by the action of $\Gau$ are physically 
indistinguishable; they are two mathematical 
representatives of the same physical state. 
Accordingly, physical observables cannot 
distinguish between these two representatives, which 
means that physical observables are constant on each 
$\Gau$-orbit in phase space. In the Hamiltonian setting 
the subset $\Gau\subset\Sym$ is usually characterised 
as the group that is generated by the constraints.
Accordingly, the space of physical observables is then 
defined to be the subset of 
phase-space functions that cannot separate points 
connected by $\Gau$, i.e. that Poisson-commute 
with the constraints on the set of points in 
phase-space allowed by the constraints. Following 
\cite{Teitelboim-YM2}, elements of $\Gau$ are 
also called ``proper gauge transformations''.

The crucial observation is that $\Sym$ is strictly 
larger than $\Gau$, so that the quotient group 
$\Asym:=\Sym/\Gau$ is again a group of symmetries, 
now to be interpreted as \emph{proper physical 
symmetries}, in the sense of mapping states and 
solutions to \emph{new}, physically different states 
and solutions. It is this quotient group that one 
should properly address as group of asymptotic 
symmetries and which should somehow contain the 
Poincar\'e group and --- possibly --- more. 
Note that $\Asym$ contains residuals of those 
``gauge transformations'' whose fall-off is too 
weak in order to be generated by constraints. 
These are often called ``improper gauge 
transformations'' \cite{Teitelboim-YM2}. 

It has long been realised the insufficient distinction  
between proper and improper gauge transformations 
may result in apparently paradoxical conclusions,
like that of an apparent violation of conservation of
global non-abelian charges which follows as consequence 
if long-ranging (and hence improper) gauge 
transformations are taken for proper ones; see, 
e.g., \cite{Schlieder:1981}.  Strictly speaking, the 
improper gauge transformations do not only contain 
those with insufficient fall off, they also may contain 
those of rapid fall-off which are not in the component 
of the identity. This is because the group $\Gau$ that 
is generated by the constraints is, by definition, 
connected. Elements outside the 
component of the identity are 
sometimes referred to as ``large gauge 
transformations''. 

Quite generally, improper gauge transformations 
will combine with other symmetries, like the 
Poincar\'e group, into the group $\Asym$. That 
combination need not be a direct product. Often 
it is a semi-direct product or, more generally, 
an extension of one group by the other. 
In fact, non-trivial extensions already appear 
when large gauge transformations are properly 
taken into account, with potentially interesting 
consequences for the physical content of the theory. 
For example, it may happen that the electromagnetic 
$\mathrm{U}(1)$ is extended to its (non-compact) 
universal cover $\mathbb{R}$, or that the spatial 
$\mathrm{SO}(3)$ is extended to its universal 
cover $\mathrm{SU}(2)$; see \cite{Giulini:1995}.

Previous studies of Yang-Mills theory in Hamiltonian 
formulation include~\cite{Teitelboim-YM1,Teitelboim-YM2}
among others. Although the focus is on the 
spherically-symmetric case, they nevertheless 
highlight some general and important features. 
We also mention the detailed discussion of boundary 
conditions allowing for globally charged states in 
\cite{Chrusciel.Kondracki:1987}.

Based on the results obtained in the study of the 
asymptotic symmetries of Yang-Mills fields at null
 infinity~\cite{Strominger-YM,Barnich-YM} and 
of the results obtained in the Hamiltonian approach 
of other gauge theories, such as 
electrodynamics~\cite{Henneaux-ED} and general
relativity~\cite{Henneaux-GR}, one expects to find a 
well-defined Hamiltonian formulation of the 
non-abelian Yang-Mills theory, which features a 
canonical action of a non-trivial group of 
asymptotic symmetries. Quite surprisingly, we 
were not able to obtain this result. Rather, we 
did find a well-defined Hamiltonian formulation 
of the theory, but the group of asymptotic symmetries 
turned out to be trivial in this case and, accordingly, 
the total colour charge had do vanish. Moreover, we find 
that if one tries to weaken the parity conditions in 
order to accommodate for a non-trivial asymptotic-symmetry 
group and for a non-vanishing value of the total 
colour charge one either has to give up the existence of 
a symplectic form or looses the Hamiltonian action 
of the Poincar\'e transformations. 

The paper is organised as follows.
In section~\ref{sec:Lagrangian-Hamiltonian-YM} we begin 
with a brief review of the Hamiltonian formulation,
thereby outlining our assumptions and also fixing the 
notation. In this introductory section, we do not 
pay much attention to typical issues of a proper
Hamiltonian formulation, such as the finiteness of the symplectic form and the functional-differentiability 
of the Hamiltonian, as they would be the subject of thorough discussions in the next sections.
Specifically, in section~\ref{sec:Lorentz-fall-off}, we infer the fall-off conditions of the fields from the requirement that they should support Poincar\'e transformations. 
In addition, in section~\ref{sec:parity}, we find parity conditions, which, in combination with the fall-off conditions, make the theory to have a finite symplectic structure, a finite and functionally differentiable Hamiltonian, and a canonical action of the Poincar\'e group. However, these parity conditions seem too 
strong in that they exclude the possibility of a 
non-trivial asymptotic Lie-algebra of symmetries 
and in preventing us to have a non-zero total colour charge. At this stage our finding is somewhat 
analogous to that in \cite{Henneaux-ED} for 
electrodynamics and not too surprising.
In section~\ref{sec:relax-parity-ED} we review how 
this issue was resolved for electrodynamics in \cite{Henneaux-ED}, which leads us to try a similar
strategy in the Yang-Mills case in 
section~\ref{sec:relax-parity-YM}. Interestingly, 
in the non-abelian case, this strategy now seems 
to manifestly fail for reasons that we outline 
in detail. Finally, our conclusions are stated in section~\ref{sec:conclusions}.

\subsection*{Conventions and notation}
\noindent
Throughout this paper, we adopt the following conventions.
Lower-case Greek indices denote spacetime components, e.g.
 $\alpha=0,1,2,3$, lower-case Latin indices denote spatial
 components, e.g. $a=1,2,3$, and lower-case barred Latin
indices denote angular components, e.g. 
$\bar a=\theta,\varphi$. We adopt the mostly-plus convention $(-,+,+,+)$ for the spacetime four-metric 
${}^4 g$. 

Moreover, upper-case latin indices denote the $\su (N)$
components and range from $1$ to $N^2-1$. We identify 
$\su (N)$ with the image of its fundamental (also called
``defining'') representation, in which its elements 
are represented by $N^2-1$ anti-hermitian $N\times N$ 
matrices $\{ T_A\}_{A=1,\dots,N^2-1}$. In this fashion, 
we embed the Lie algebra into the associative algebra 
of endomorphisms with (associative) product being matrix 
multiplication. In this way, the Lie product becomes the 
associative product's commutator and, moreover, we may 
speak of (associative) products of elements of the Lie 
algebra, like, e.g., in formulae \eqref{eq:DefKillingFormFundRep} 
and \eqref{eq:CompModifiedKillingProduct} below, which is 
very useful --- though not necessary --- for many later 
calculations and which would not make sense on an 
abstract level of Lie algebras. Note that the matrix
product of elements in $\su (N)$ will generally yield 
matrices outside $\su (N)$.

The structure constants $f^A{}_{BC}$ are defined by the 
relation $[T_B,T_C]=f^A{}_{BC} T_A$. On $\su(N)$, we 
consider a positive-definite inner product, which we 
obtain from the Killing form, $\kappa$, through 
multiplication with $(-2N)^{-1}$. This will turn out
to be a convenient normalisation in later calculations. To explain this in slightly 
more detail, we recall that the Killing form itself is a symmetric bilinear form 
on the Lie algebra, defined by 
\begin{equation}
 \label{eq:DefKillingForm}
\kappa(T_A,T_B):=
\tr\bigl(\ad_{T_A}\circ \ad_{T_B}\bigr)
=f^N{}_{AM}f^M{}_{BN}\,,
\end{equation}
where $\circ$ denotes the operation of composition 
(of endomorphisms). On $\su(N)$ the Killing form 
defines a negative-definite inner product (like for 
any compact Lie algebra).
Moreover, through our identification of $\su(N)$ with 
its image under the fundamental representation, we 
can eliminate the occurrence of the adjoint representation 
in the definition of the inner product and express it
directly trough traces of products of Lie algebra
elements in a form that is only valid for $\su(N)$: 
\begin{equation}
 \label{eq:DefKillingFormFundRep}
\kappa(T_A,T_B)=
2N\,\tr(T_AT_B)\,.
\end{equation}
Here, juxtaposition of matrices in $\su(N)$ refers 
to matrix multiplication. Now, the inner product 
we shall be using is 
\begin{equation}
 \label{eq:DefModifiedKillingForm}
S:=-\frac{1}{2N}\, \kappa\,.
\end{equation}
Its components with respect to the basis 
$\{ T_A\}_{A=1,\dots,N^2-1}$ are therefore
\begin{equation}
 \label{eq:CompModifiedKillingProduct}
S_{AB}= - \tr (T_A T_B)\,.
\end{equation}
Its inverse has components $S^{AB}$ 
and satisfies 
\begin{equation}
 \label{eq:InvModifiedKillingProduct}
S^{AM}S_{BM}=\delta^A_B\,.
\end{equation}
In our paper, we shall exclusively use $S$ and hence 
continue, for simplicity, to refer to it as ``Killing
inner product'', keeping in mind that it is actually 
a negative multiple of $\kappa$.
  
We use $S_{AB}$ and $S^{AB}$ to raise 
and lower indices in the standard 
fashion, e.g., in order to define the 
index-lowered structure constants
\begin{equation}
\label{eq:DefIndexLoweredSC}
 f_{ABC} \eqdef S_{AA'} f^{A'}{}_{BC}\,,
\end{equation}
 which are easily seen to be completely antisymmetric, using $f_{ABC}=-\tr\bigl(T_A[T_B,T_C]\bigr)$ and the cyclicity 
of the trace.

Finally, given two Lie-algebra-valued functions 
$\phi\eqdef \phi^A T_A$ and $\psi\eqdef \psi^A T_A$, 
we denote their positive-definite inner product
by a dot, like
\begin{equation}
\label{eq:DefSprod}
 \sprod{\phi}{\psi} \eqdef \phi^A S_{AB} \psi^B\,,
\end{equation}
and the commutators by 
\begin{equation}
\label{eq:DefExtprod}	
 \extprod{\phi}{\psi} \eqdef [\phi,\psi] \,.
\end{equation}
Inner product and commutator then obey the 
familiar rule
\begin{equation}
\label{eq:SprodExtprod}	
 \sprod{\phi}{(\extprod{\psi}{\chi})}=
 \sprod{\psi}{(\extprod{\chi}{\phi})}=
 \sprod{\chi}{(\extprod{\phi}{\psi})} \,,
\end{equation}
with the same cyclic property of the triple product.
In this notation, the Jacobi identity reads
\begin{equation}
\label{eq:JacobiIdenity}	
 \extprod{\phi}{(\extprod{\psi}{\chi})}+
 \extprod{\psi}{(\extprod{\chi}{\phi})}+
 \extprod{\chi}{(\extprod{\phi}{\psi})}=0 \,.
\end{equation}
In addition, by means of the positive-definite 
inner product, we may and will identify (as vector 
spaces) the Lie-algebra and its dual and this we 
extend to functions. So, if $\hat\phi$ is 
dual-Lie-algebra-valued function, we assign it to the 
unique Lie-algebra-valued function $\phi$ 
satisfying $\hat\phi(\psi)=\sprod{\phi}{\psi}$ 
for all $\psi$. Examples of such dual-Lie-algebra-valued 
functions that we will encounter in the 
following sections and identify with their 
corresponding Lie-algebra-valued functions
are the conjugated momenta $\pi^\alpha$ and 
the Gauss constraint $\mathscr{G}$.

\section{Hamiltonian of free Yang-Mills theory} 
\label{sec:Lagrangian-Hamiltonian-YM}
In this section, we briefly review the Hamiltonian formulation of Yang-Mills theory on a flat Minkowski background.
We follow mostly the line of argument and the notation of~\cite{Dirac-book}, which discusses the case of electrodynamics.
In order to have a description as self contained as possible, we begin by deriving the Hamiltonian of free Yang-Mills theory from the more-commonly-used Lagrangian picture, in which the action is
\begin{equation} \label{action-Lagrangian}
 S[A_\alpha,\dot A_\alpha;g] = -\frac{1}{4} \int d^4 x \sqrt{-{}^4g} \, {}^4g^{\alpha \gamma} \, {}^4g^{\beta \delta} \,  \sprod{F_{\alpha \beta}}{F_{\gamma \delta}}
 +(\text{boundary})\,,
\end{equation}
where $A_\alpha$ is the $\su(N)$-valued one-form potential,  $F_{\alpha \beta} \eqdef \partial_\alpha A_\beta -\partial_\beta A_\alpha + \extprod{A_\alpha}{A_\beta}$ is the curvature two-form, and ${}^4 g$ is the four-dimensional flat spacetime metric.
The boundary term in the action is necessary to make the Lagrangian functionally-differentiable and to make the following manipulations meaningful.
For now, we just assume its existence and postpone a thorough discussion about it to the next sections.

The spacetime four-metric ${}^4g$ is $(3+1)$-decomposed into
\[
 {}^4g_{\alpha \beta}=
 \left(
 \begin{array}{c|c}
  -1	& 0	\\ \hline
  0	& g_{ab}
 \end{array}
 \right) \,.
\]
Although we are dealing with flat Minkowski spacetime, it is more convenient to leave the three-metric $g$ in general coordinates for now.
Later on, we will express it in radial-angular coordinates, but there is no advantage in doing it at this stage.
From now on, spatial indices are lowered and raised using the three-metric $g$ and its inverse.
The action becomes $S=\int dt L[A,\dot A;g]$, where the Lagrangian is
\begin{equation} \label{Lagrangian}
  L[A_\alpha,\dot A_\alpha;g] =\int d^3 x \sqrt{g} \left[
  \frac{1}{2} g^{ab} \sprod{F_{0a}}{F_{0b}}
  -\frac{1}{4} \sprod{F_{ab} }{F^{ab}}
  \right] + (\text{boundary}) \,.
\end{equation}
The variation of the Lagrangian~(\ref{Lagrangian}) with respect to $\dot A_\alpha$ yields the conjugated three-momenta
\begin{equation} \label{momenta}
 \pi^a \eqdef \frac{\delta L}{\delta \dot A_a} =\sqrt{g} \, g^{ab} F_{0b} \,,
\end{equation}
which are vector densities of weight $+1$, and the primary constraints
\begin{equation} \label{primary-constraints}
 \pi^0 \eqdef \frac{\delta L}{\delta \dot A_0} \weq 0 \,.
\end{equation}
Note that these are $N^2-1$ independent constraints since $\pi^0$ has $N^2-1$ independent components. 
From this, one obtains straightforwardly the Hamiltonian
\begin{equation} 
\label{Hamiltonian_0}
\begin{split}
 H_0[A,\pi;g;\mu]=
 & \int d^3 x \left[ 
 \frac{\sprod{\pi^a}{\pi_a}}{2\sqrt{g}} 
 +\frac{\sqrt{g}}{4} \sprod{F_{ab}}{F^{ab}}
 -\sprod{A_0}{(\partial_a \pi^a+\extprod{A_a}{\pi^a})}
 +\sprod{\mu}{\pi^0}
 \right]\\
 &+ (\text{boundary}) \,,
\end{split}
\end{equation}
after using the definition $H\eqdef \int d^3 x \, \sprod{\pi^\alpha}{\dot A_\alpha}-L$, replacing $\dot A_a$ with $\pi^a$ by means of~(\ref{momenta}), adding the constraints~(\ref{primary-constraints}) with a Lagrange multiplier $\mu$, and absorbing $\dot A_0$ in the Lagrange multiplier $\mu$.
Finally, the symplectic form, from which the Poisson brackets ensue, is
\begin{equation} \label{symplectic-form_0}
 \Omega_0[A_\alpha,\pi^\alpha]=
 \int d^3 x \, \sprod{\extder \pi^\alpha \wedge}{\extder A_\alpha} \eqdef
 \int d^3 x \, \extder \pi^\alpha_A \wedge \extder A_\alpha^A \,,
\end{equation}
where the bold $\extder$ and $\wedge$ are, respectively, the exterior derivative and the wedge product in phase space.
Moreover, the symbol $\sprod{\wedge }{}$ means that, at the same time, we are doing the wedge product in phase space and (the negative of) the Killing inner product in the $\su(N)$ degrees of freedom.

\subsection{Secondary constraints and constraints' algebra}
The constraints $\pi^0 \weq 0$ are not preserved by time evolution.
Indeed,
\begin{equation}
 \dot \pi^0 = \{ \pi^0, H_0 \} = \partial_a \pi^a + \extprod{A_a}{\pi^a} \,,
\end{equation}
which is, in general, different from zero.
Therefore, one enforces the secondary constraints
\begin{equation} \label{secondary-constraints}
 \mathscr{G} \eqdef \partial_a \pi^a + \extprod{A_a}{\pi^a} \weq 0 \,,
\end{equation}
so that the primary constraints~(\ref{primary-constraints}) are preserved by time evolution.
Note that the expression in~(\ref{secondary-constraints}) is precisely the term multiplied by $A_0$ in the Hamiltonian~(\ref{Hamiltonian_0}) and that it is build using the gauge-covariant derivative $D_b \pi^a \eqdef \partial_b \pi^a + \extprod{A_b}{\pi^a}$.

At this point, one needs to ensure that also the secondary constraints~(\ref{secondary-constraints}) are preserved by time evolution.
This is indeed the case since
\begin{equation}
 \dot{\mathscr{G}} = \{ \mathscr{G}, H_0 \} = -\extprod{A_0}{\mathscr{G}} \weq 0 \,.
\end{equation}
This shows that we have found all the constraints of the theory, $\pi^0$ and $\mathscr{G}$.
These constraints are first class.
Indeed, if we decompose them into components, $\pi^0_A \eqdef \sprod{\pi^0}{T_A}$ and
\mbox{$\mathscr{G}_A \eqdef \sprod{\mathscr{G}}{T_A}$}, and we compute their Poisson brackets, we get
\begin{equation}
\label{constraint-algebra}
\begin{split}
 \{ \pi^0_A (x), \pi^0_B (x') \}&=0 \,, \\
 \{ \pi^0_A (x), \mathscr{G}_B (x') \}&=0 \,,\\ 
 \{ \mathscr{G}_A (x), \mathscr{G}_B (x') \}&=f^M{}_{AB} \, \mathscr{G}_M (x) \delta (x-x') \,.
\end{split}
\end{equation}
Notably, the last one of the expressions above shows that the constraints  $\{ \mathscr{G}_A\}_{A=1,\dots,N^2-1}$ form a Poisson-representation of the $\su (N)$ algebra.

\subsection{Hamiltonian of free Yang-Mills theory}
As well as the primary constraints~(\ref{primary-constraints}), also the secondary constraints~(\ref{secondary-constraints}) need to be included in the  Hamiltonian~(\ref{Hamiltonian_0}) multiplied by a Lagrange multiplier $\lambda$.
Doing so and reabsorbing $A_0$ in the definition of $\lambda$, one obtains the extended Hamiltonian of free Yang-Mills theory
\begin{equation} 
\label{extended-Hamiltonian}
\begin{split} 
H_{\text{ext}}[A_\alpha,\pi^\alpha;g;\mu,\lambda]=
 \int  d^3 x & \left[ 
 \frac{\sprod{\pi^a}{\pi_a}}{2\sqrt{g}} 
 +\frac{\sqrt{g}}{4} \sprod{F_{ab}}{F^{ab}}
 +\sprod{\mu}{\pi^0}
 +\sprod{\lambda}{\mathscr{G}}
 \right]\\
 +&(\text{boundary})\,.
\end{split}
\end{equation}
As in the case of electrodynamics, one can remove the degrees of freedom corresponding to $\pi^0$ and $A_0$, since they do not contain any physical information.
Indeed, their equations of motion are
\begin{align} \label{eom-zero-comp}
 \dot A_0=\mu \,, &&
 \dot \pi^0=0 \,, &&
 \pi^0 \weq 0 \,,
\end{align}
so that the derivative of $A_0$ is completely arbitrary and $\pi^0$ is identically zero.
Therefore, we discard these degrees of freedom obtaining the symplectic form
\begin{equation} 
\label{symplectic-form}
 \Omega[A,\pi]=\int d^3 x \, \sprod{\extder \pi^a \wedge}{\extder A_a}
\end{equation}
and the Hamiltonian of free Yang-Mills theory
\begin{equation} 
\label{Hamiltonian}
 H[A,\pi;g;\lambda]=
 \int d^3 x \left[ 
 \frac{\sprod{\pi^a}{\pi_a}}{2\sqrt{g}} 
 +\frac{\sqrt{g}}{4} \sprod{F_{ab}}{F^{ab}}
 +\sprod{\lambda}{\mathscr{G}}
 \right] +(\text{boundary}) \,,
\end{equation}
where the only constraints left are the $(N^2-1)$ first-class Gauss-like constraints
\begin{equation} \label{gauss-constraint}
 \mathscr{G} \eqdef 
 \partial_a \pi^a  + \extprod{A_a}{\pi^a}
 =D_a \pi^a
 \weq 0\,.
\end{equation}

Finally, the knowledge of the symplectic form~(\ref{symplectic-form}) and of the Hamiltonian~(\ref{Hamiltonian}) allows one to compute the equations of motion
\begin{align} \label{eoms-YM1}
 \dot{A}_a &= \{ A_a , H \}
 =\frac{\pi_a}{\sqrt{g}}
 % - \partial_a \lambda +\extprod{\lambda}{A_a}
 - D_a \lambda
 \,,\\
 \label{eoms-YM2}
 \dot{\pi}^a &= \{ \pi^a , H \}
 =\partial_b (\sqrt{g}\, F^{ba}) +\sqrt{g}\, \extprod{A_b}{F^{ba}}+\extprod{\lambda}{\pi^a} \,.
\end{align}
The presence of the Gauss constraints~(\ref{gauss-constraint}) in the Hamiltonian~(\ref{Hamiltonian}) causes the equations of motion above to include a gauge transformation, whose gauge parameter is the arbitrary function $\lambda (x)$.
We briefly discuss gauge transformations in the next subsection.

\subsection{Gauge transformations} \label{subsec:gauge-transformations-intro}
Gauge transformations are those transformations generated by first-class constraints, like the ones that we have encountered so far in this paper.
In particular, the canonical generator of the gauge transformations of Yang-Mills is
\begin{equation} \label{gauge-generator}
 G[\lambda] \eqdef \int d^3 x\, \sprod{\lambda (x)}{\mathscr{G} (x)} \,,
\end{equation}
which is the Gauss constraints~(\ref{gauss-constraint}) smeared with an arbitrary function $\lambda (x)$.
The above expression is precisely the last term appearing in the Hamiltonian~(\ref{Hamiltonian}).
The variation of the gauge generator~(\ref{gauge-generator}) is
\begin{equation} \label{gauge-generator-variation}
 \delta G[\lambda] = \int d^3 x\, \Big[
 \sprod{-\delta \pi^a}{(\partial_a \lambda+\extprod{A_a}{\lambda})}
 -\sprod{\delta A_a}{\extprod{\lambda}{A_a}}
 \Big]+
 \oint_{S^2_{\infty}} d^2 s_k \, \sprod{\lambda}{\pi^k} \,,
\end{equation}
where the last integral in the expression above has to be understood as the integral over a sphere whose radius is sent to infinity.
When the surface term in the expression above vanishes, the generator~(\ref{gauge-generator}) is functionally differentiable with respect to the canonical fields and one gets the infinitesimal gauge transformations
\begin{align} \label{gauge-transformations1}
 \delta_\lambda A_a &\eqdef \{ A_a, G[\lambda] \} = -D_a \lambda \,, \\
 \label{gauge-transformations2}
 \delta_\lambda \pi^a &\eqdef \{ \pi^a, G[\lambda] \} = \extprod{\lambda}{\pi^a} \,,
\end{align}
which are exactly the last terms appearing in~(\ref{eoms-YM1}) and in~(\ref{eoms-YM2}).
As it is well known, two field configurations related by gauge transformations are physically equivalent and the degrees of freedom in the description of the theory are redundant.
The infinitesimal transformations above can be integrated to get the gauge transformations with parameter $\mathcal{U}\eqdef \exp (-\lambda) \in \SU(N)$
\begin{align} \label{gauge-transformations-full1}
 A_a \mapsto \Gamma_\mathcal{U} (A_a) & = \mathcal{U}^{-1} A_a \,\mathcal{U}+\mathcal{U}^{-1} \partial_a  \,\mathcal{U} \,, \\
 \label{gauge-transformations-full2}
 \pi^a \mapsto \Gamma_\mathcal{U} (\pi^a)  &= \mathcal{U}^{-1} \pi^a \,\mathcal{U} \,,
\end{align}
where the products on the right-hand sides are products among matrices.

Whether or not the surface term in~(\ref{gauge-generator-variation}) is zero depends on the asymptotic behaviour of the canonical fields and of the gauge parameter $\lambda(x)$, which topic is going to be thoroughly discussed in the following sections.
After this discussion is made, we will come back to gauge transformations and examine them in more detail in section~\ref{subsec:gauge-transformations}.

This concludes the brief survey of the derivation of the Yang-Mills free Hamiltonian.
The symplectic form~(\ref{symplectic-form}), the Hamiltonian~(\ref{Hamiltonian}), and the Gauss constraints~(\ref{gauss-constraint}) are the starting points and the fundamental parts of the ensuing discussion, whose goal is to provide a well-defined Hamiltonian formulation of the Yang-Mills theory.

\section{Poincar\'e transformations and fall-off conditions} \label{sec:Lorentz-fall-off}
The symplectic form and the Hamiltonian derived at the end of the last section are not yet providing a well-defined Hamiltonian description of free Yang-Mills on a Minkowski spacetime.
This happens mostly for two reasons, which were left aside in the previous section.
First, the integral in~(\ref{symplectic-form}) might not be finite and, as a consequence, the symplectic form would not be well defined.
Secondly, one needs to make sure that also the Hamiltonian~(\ref{Hamiltonian}) is finite and, moreover, functionally differentiable with respect to the canonical fields.
In order to achieve this, it may happen that one needs to add a boundary term in the Hamiltonian.
In addition to these two problems, we would also like to include a well-defined canonical action of the Poincar\'{e} group on the fields.

The method to solve the aforementioned problems works as follows.
First, one makes the space of allowed field configurations smaller by requiring that the fields satisfy some fall-off conditions at spatial infinity.
This step will be the topic of this section.
The fall-off conditions should be strong enough, so that the Hamiltonian is finite and the symplectic form is, at most, logarithmically divergent.
At the same time, they should be weak enough not to exclude any potentially interesting solution of the equations of motion.
Moreover, since one wishes to include the Poincar\'{e} transformations as symmetries of the theory, one also needs to impose that the fall-off conditions are preserved by Poincar\'{e} transformations.
For, otherwise, the transformations would map allowed filed configurations to non-allowed ones.

Second, one makes the symplectic form finite by requiring that the leading terms in the asymptotic expansion of the fields have a definite parity, either even or odd, as functions on the sphere.
These parity conditions are chosen so that the logarithmically divergent contribution to the symplectic form is, in fact, zero.
In some cases, such as electrodynamics, it is also possible to relax a bit the parity conditions~\cite{Henneaux-ED}, so that one makes the space of allowed field configurations bigger.
We will discuss parity conditions and their possible relaxation in sections~\ref{sec:parity},~\ref{sec:relax-parity-ED}, and~\ref{sec:relax-parity-YM}.

The reason for leaving the symplectic form logarithmically divergent when imposing the fall-off conditions and making it finite with parity conditions, rather than making it finite directly by means of the fall-off conditions, is that, in this way, the phase space is larger and, therefore, one obtains potentially more solutions of the equations of motion.

\subsection{Poincar\'e transformations of the fields} \label{subsec:poincare-transformations}
In this subsection, we determine how the fields transform under Poincar\'{e} transformations.
We begin by establishing the transformation of the fields under a generic hypersurface deformation.
Then, we specialize the results in the case of a deformation corresponding to Poincar\'e transformations.

A generic hypersurface deformation can be decomposed into a component normal to the hypersurface, which we denote by $\xi^\perp$, and components tangential to the hypersurface, denoted by $\xi^i$.
The transformation of the fields under such a deformation is generated by
\begin{equation} \label{diffeo-generator}
 H[\xi^\perp,\xi^i]=
 \int d^3 x \, \big[
 \xi^\perp \, \mathscr{H} (A,\pi;g)
 + \xi^i \, \mathscr{H}_i (A,\pi;g)
 \big]
 +(\text{boundary}) \,.
\end{equation}
Whether or not the generator~(\ref{diffeo-generator}) is finite and functionally differentiable depends on the asymptotic behaviour of $\xi$ and of the canonical fields.
At the moment, we assume that~(\ref{diffeo-generator}) is finite and functionally differentiable and we check \textit{a posteriori} in section~\ref{sec:parity} if this is true for the Poincar\'e transformations, after we have specified the fall-off and parity conditions of the canonical fields.

One way to obtain an explicit expression for~(\ref{diffeo-generator}) would consist in redoing the analysis of section~\ref{sec:Lagrangian-Hamiltonian-YM} using a general $(3+1)$-decomposition for the metric, which includes lapse and shift.
The Hamiltonian that one would find would correspond to the generator~(\ref{diffeo-generator}), after identifying $\xi^\perp$ with the lapse and $\xi^i$ with the shift.
Another and quicker way, which provides the same result, consists in noting that the generator~(\ref{diffeo-generator}) needs to produce a time translation when $\xi^\perp=1$ and $\xi^i=0$.
Therefore, in this case, it should coincide with the Hamiltonian~(\ref{Hamiltonian}), from which one reads
\begin{equation} \label{super-Hamiltonian}
 \mathscr{H} = \frac{\sprod{\pi^a}{\pi_a}}{2\sqrt{g}} + \frac{\sqrt{g}}{4} \sprod{F_{ab}}{F^{ab}} + \sprod{\lambda}{\mathscr{G}} \,. 
\end{equation}
Note that, due to the last term in~(\ref{super-Hamiltonian}), the generator~(\ref{diffeo-generator}) includes a gauge transformation with gauge parameter $\zeta \eqdef \xi^\perp \lambda$.
The tangential part of the generator $\mathscr{H}_i$ can be determined by geometrical reasons.
One simply requires that $A_a$ behaves like a covector field and $\pi^a$ like a density-one vector field under tangential deformations.
As a results, one finds
\begin{equation} \label{super-momentum}
 \mathscr{H}_i = \sprod{\pi^a}{\partial_i A_a} - \partial_a (\sprod{\pi^a}{A_i}) \,.
\end{equation}
Having determined completely the form of the generator~(\ref{diffeo-generator}), one can compute the transformation of the fields under a generic hypersurface deformation, finding
\begin{align} \label{poincare-transformations1}
 \delta_{\xi,\zeta} A_a &\eqdef \big\{ A_a, H[\xi^\perp,\xi^i] \big\}
 =\xi^\perp \frac{\pi_a}{\sqrt{g}}+\xi^i \partial_i A_a+\partial_a \xi^i A_i-D_a \zeta \,,	\\
 %%%%
 \label{poincare-transformations2}
 \delta_{\xi,\zeta} \pi^a &\eqdef \big\{ \pi^a, H[\xi^\perp,\xi^i] \big\}
 =%\partial_b(\xi^\perp \sqrt{g}F^{ba})+\xi^\perp \sqrt{g} \extprod{A_b}{F^{ba}}
  \sqrt{g} \, D_b(\xi^\perp F^{ba})
 +\partial_i (\xi^i \pi^a) - \partial_i \xi^a \pi^i+\extprod{\zeta}{\pi^a} \,.
\end{align}

Finally, one  can find out the behaviour of the canonical fields under Poincar\'e transformations.
Indeed, in Cartesian coordinates $(t,x^i)$, these corresponds to a hypersurface deformation parametrized by
\begin{align} \label{poincare-xi-cartesian}
 \xi^\perp =a^\perp + b_i x^i
 \qquad \text{and} \qquad
 \xi^i= a^i +\omega^{i}{}_j x^j \,,
\end{align}
where $a^\perp$ is responsible for the time translation, $a^i$ for the spatial translations, $b^i$ for the Lorentz boost, and the antisymmetric $\omega_{ij} \eqdef g_{i\ell}\omega^{\ell}{}_j$ for the spatial rotations.
Note that, following~\cite{Henneaux-GR,Henneaux-ED}, we have absorbed the contribution  of the boost $t \,b^i$, which would appear in $\xi^i$, into the parameters $a^i$.
The reason for doing so is that these two terms have the same dependence on the radial distance in the asymptotic expansion at spatial infinity.

For the following discussion, it is actually more convenient to move to spherical coordinates $(t,r,\barr{x})$, where $\barr{x}$ are coordinates on the unit two-sphere, such as the usual $\theta$ and $\varphi$.
The flat three-metric is
\[
 g_{ab}=
 \left(
 \begin{array}{c|c}
  1	& 0	\\
  \hline
  0	& r^2 \, \barr{\gamma}_{\bar a \bar b}
 \end{array}
 \right) \,,
\]
where $\barr{\gamma}_{\bar a \bar b}$ is the metric of the unit round sphere and indices with bars above, such as $\bar a$, run over the angular components.
Using these coordinates, the components of the vector field~(\ref{poincare-xi-cartesian}) corresponding to Poincar\'{e} transformations are
\begin{align} \label{poincare-xi}
 \xi^{\perp}=r b +T \,, &&
 \xi^r = W \,, &&
 \xi^{\bar a} = Y^{\bar a} + \frac{1}{r} \barr{\gamma}^{\bar a \bar m}\, \partial_{\bar m} W \,.
\end{align}
In the above expression, $b$, $Y^{\bar a}$, $T$, and $W$ are functions on the sphere satisfying the equations 
\begin{align}
 \barr{\nabla}_{\bar a} \barr{\nabla}_{\bar b} W +\barr{\gamma}_{\bar a \bar b} W=0 \,, &&
 \barr{\nabla}_{\bar a} \barr{\nabla}_{\bar b} b +\barr{\gamma}_{\bar a \bar b} b=0 \,, &&
 \mathcal{L}_Y \barr{\gamma}_{\bar a \bar b}=0 \,, &&
 \partial_{\bar a} T=0 \,,
\end{align}
where $\barr{\nabla}$ is the covariant derivative on the unit round sphere.
Moreover, $b$, $Y^{\bar a}$, $T$, and $W$ are related to the parameters $a^\perp$, $a^i$, $m^i \eqdef - \epsilon^{i j k} \omega_{jk} /2$, and $b^i$ by the expressions
\begin{align}
 b(\theta,\varphi) &=b_1 \sin \theta \cos \varphi +b_2 \sin \theta \sin \varphi+b_3 \cos\theta \,,\\
 Y(\theta,\varphi) &=
 m_1 \left( -\sin \varphi \frac{\partial}{\partial \theta} - \frac{\cos \theta}{\sin \theta} \cos \varphi \frac{\partial}{\partial \varphi} \right)\nonumber\\
 &+m_2 \left( \cos \varphi \frac{\partial}{\partial \theta} - \frac{\cos \theta}{\sin \theta} \sin \varphi \frac{\partial}{\partial \varphi} \right)\nonumber\\
 &+m_3 \frac{\partial}{\partial \varphi} \,,\\
 W(\theta,\varphi) &=a_1 \sin \theta \cos \varphi +a_2 \sin \theta \sin \varphi+a_3 \cos\theta \,, \\
 T(\theta,\varphi) &=a^\perp \,,
\end{align}
where we have used explicitly the usual $\theta$ and $\varphi$ as angular coordinates.

The Poincar\'e transformations of the fields are, therefore, obtained by inserting~(\ref{poincare-xi}) into the expressions~(\ref{poincare-transformations1}) and~(\ref{poincare-transformations2}).
There is no need to write down the explicit expression of the Poincar\'e transformations at this stage.
We will show explicitly how the transformations act on the asymptotic part of the fields after we have determined the fall-off behaviour of the fields.

\subsection{Fall-off conditions of the fields}
In this subsection, we determine the fall-off conditions of the fields.
In order to do this, we demand the following requirements to be satisfied.
First, the symplectic form~(\ref{symplectic-form}) should be, at most, logarithmically divergent.
Second, the fall-off conditions of the fields should be preserved by the Poincar\'e transformations, discussed in the last subsection.
Third, the asymptotic expansion of the fields should be of the form
\begin{equation} 
\label{fall-off-conditions-preliminary}
\begin{aligned}
  A_r (r,\barr{x}) &= \frac{1}{r^\alpha} \, \barr{A}_r (\barr{x}) +\bigo(1/r^{\alpha+1}) \,,
  & \pi^r (r,\barr{x}) &= \frac{1}{r^{\alpha'}} \, \barr{\pi}^r (\barr{x}) +\bigo(1/r^{\alpha'+1})\,, \\
  A_{\bar{a}} (r,\barr{x}) &= \frac{1}{r^\beta} \, \barr{A}_{\bar{a}} (\barr{x}) +\bigo(1/r^{\beta+1}) \,, 
  & \pi^{\bar{a}} (r,\barr{x}) &= \frac{1}{r^{\beta'}} \, \barr{\pi}^{\bar{a}} (\barr{x}) + \bigo(1/r^{\beta'+1}) \,. \\
\end{aligned}
\end{equation}
The dependence of the fields on the time coordinate $t$, though present, is not denoted explicitly in the above expressions and in the following ones. 
Note that we require the leading term in the expansion to be an integer power of $r$ and the first subleading term in the expansion to be the power of $r$ with exponent reduced by one.
Functions whose fall-off behaviour is between the two next powers of $r$, such as those one could build using logarithms, are excluded at the first subleading order.
Fourth, the fall-off conditions should be the most general ones compatible with the previous three requirements, so that the space of allowed field configurations is as big as possible.
In addition, we also expand the gauge parameter appearing in~(\ref{poincare-transformations1}) and~(\ref{poincare-transformations2}) according to
\begin{equation} \label{fall-off-conditions-gauge-preliminary}
 \zeta= \frac{1}{r^\delta} \barr{\zeta} (\barr{x}) +\bigo(1/r^{\delta+1}) \,.
\end{equation}

To begin with, the requirement that the symplectic form~(\ref{symplectic-form}) is, at most, logarithmically divergent implies the relations
\begin{equation} \label{fall-off-from-symplectic}
 \alpha+\alpha' \ge 1
 \qquad \text{and} \qquad
 \beta+\beta' \ge 1
\end{equation}
among the exponents defined in~(\ref{fall-off-conditions-preliminary}).
If the two inequalities above are satisfied strictly, then the symplectic form is actually finite.

Then, one checks when the fall-off conditions~(\ref{fall-off-conditions-preliminary}) and~(\ref{fall-off-conditions-gauge-preliminary}) are preserved by the Poincar\'e transformations.
To do so, one considers the transformation of the fields, which are obtained by the combination of~(\ref{poincare-transformations1}) and~(\ref{poincare-transformations2}) with~(\ref{poincare-xi}), and inserts, into these expressions, the asymptotic expansions~(\ref{fall-off-conditions-preliminary}) and~(\ref{fall-off-conditions-gauge-preliminary}).
As a result, one finds that the fall-off conditions are preserved by the Poincar\'e transformations if
\begin{align} \label{fall-off-from-poincare}
 1 \le \alpha < 2 \,, &&
 \alpha' = \alpha -1 \,, &&
 \beta=0 \,, &&
 \beta'=1 \,, &&
 \delta \ge 0 \,.
\end{align}
Note that these equations already imply~(\ref{fall-off-from-symplectic}).
Finally, requiring that the fall-off conditions are the most general ones of all the possible ones, one obtains that the fields behave asymptotically as
\begin{equation} \label{fall-off-conditions}
\begin{aligned}
  A_r (r,\barr{x}) &= \frac{1}{r} \barr{A}_r (\barr{x}) +\bigo(1/r^2) \,,
  & \pi^r (r,\barr{x}) &= \barr{\pi}^r (\barr{x}) +\bigo(1/r)\,, \\
  A_{\bar{a}} (r,\barr{x}) &=\barr{A}_{\bar{a}} (\barr{x}) +\bigo(1/r) \,, 
  & \pi^{\bar{a}} (r,\barr{x}) &= \frac{1}{r} \barr{\pi}^{\bar{a}} (\barr{x}) + \bigo(1/r^2)
\end{aligned}
\end{equation}
and the gauge parameter behaves as
\begin{equation} \label{fall-off-conditions-gauge}
 \zeta (r,\barr{x})= \barr{\zeta} (\barr{x}) +\bigo(1/r) \,.
\end{equation}
Of course, the gauge parameter $\lambda$ appearing in~(\ref{Hamiltonian}) and~(\ref{gauge-generator}) needs to satisfy the same fall-off behaviour of $\zeta$, so that gauge transformations~(\ref{gauge-transformations1}) and~(\ref{gauge-transformations2}) preserve the fall-off conditions~(\ref{fall-off-conditions}) of the canonical fields.

To sum up,  we have determined the most general fall-off conditions of the fields and of the gauge parameter, under the requirements that they are preserved by the Poincar\'e transformations and that they make the symplectic form, at most, logarithmically divergent.
Specifically, the fall-off conditions~(\ref{fall-off-conditions}) imply that the symplectic form is precisely logarithmically divergent and not yet finite.
We will solve this issue in section~\ref{sec:parity} by means of parity conditions.
But before we do that, we spend the remainder of this section to work out the explicit expressions for the Poincar\'e transformations of the asymptotic part of the fields.

\subsection{Asymptotic Poincar\'e transformations}
The Poincar\'e transformations of the fields were not written explicitly, when they were discussed in subsection~\ref{subsec:poincare-transformations}.
We will now fix this lack, at least for what concerns the action of the Poincar\'e transformations on the asymptotic part of the fields.
The results of this subsection will be used when discussing the parity conditions in the next section.

The procedure to obtain the Poincar\'e transformation of the asymptotic part of the fields is straightforward, although a little cumbersome.
One inserts the asymptotic expansions~(\ref{fall-off-conditions}) and~(\ref{fall-off-conditions-gauge}) into the transformations~(\ref{poincare-transformations1}) and~(\ref{poincare-transformations2}) combined with~(\ref{poincare-xi}).
After neglecting all the subleading contributions, one finds
\begin{align}
 %%%% A_r
 \label{poincare-asymptotic1}
 \delta_{\xi,\zeta} \barr{A}_r =&
 \frac{b\, \barr{\pi}^r}{\sqrt{\barr{\gamma}}}
 + Y^{\bar m} \partial_{\bar m} \barr{A}_r
 + \extprod{\barr{\zeta}}{\barr{A}_r} \,,\\
 %%%% A_a
 \label{poincare-asymptotic2}
 \delta_{\xi,\zeta} \barr{A}_{\bar a} =&
 \frac{b \, \barr{\pi}_{\bar a}}{\sqrt{\barr{\gamma}}}
 +Y^{\bar m} \partial_{\bar m} \barr{A}_{\bar a} + \partial_{\bar a} Y^{\bar m} \barr{A}_{\bar m}
 %-\partial_{\bar a} \barr{\zeta} + \extprod{\barr{\zeta}}{\barr{A}_{\bar a}}
 -\barr{D}_{\bar a} \barr{\zeta}
 \,, \\
 %%%% \pi_r
 \label{poincare-asymptotic3}
 \delta_{\xi,\zeta} \barr{\pi}^r =&
 %\partial_{\bar m} \big[b \sqrt{\barr{\gamma}}\, \barr{\gamma}^{\bar m \bar n} (\partial_{\bar n} \barr{A}_r +\extprod{\barr{A}_{\bar n}}{\barr{A}_r}) \big]
 \barr{D}^{\bar m} \big( b\, \sqrt{\barr{\gamma}}\, \barr{D}_{\bar m} \barr{A}_r \big)
 %+b \sqrt{\barr{\gamma}}\, \barr{\gamma}^{\bar m \bar n} \extprod{\barr{A}_{\bar m}}{(\partial_{\bar n} \barr{A}_r +\extprod{\barr{A}_{\bar n}}{\barr{A}_r})}+\\
 %&
 +\partial_{\bar m} (Y^{\bar m} \barr{\pi}^r)
 +\extprod{\barr{\zeta}}{\barr{\pi}^r} \,, \\
 %%%% \pi_a
 \label{poincare-asymptotic4}
 \delta_{\xi,\zeta} \barr{\pi}^{\bar a} =& 
 %\partial _{\bar m} \big[ b \sqrt{\barr{\gamma}}\, \barr{\gamma}^{\bar a \bar s}\, \barr{\gamma}^{\bar m \bar n} (\partial_{\bar n} \barr{A}_{\bar s}-\partial_{\bar s} \barr{A}_{\bar n}+\extprod{\barr{A}_{\bar n}}{\barr{A}_{\bar s}}) \big]
 \barr{D}_{\bar m} \big( b \, \sqrt{\barr{\gamma}}\, \barr{F}^{\bar m \bar a})
 %-b \sqrt{\barr{\gamma}} \, \barr{\gamma}^{\bar a \bar m} \, \extprod{\barr{A}_r}{(\partial_{\bar m} \barr{A}_r +\extprod{\barr{A}_{\bar m}}{\barr{A}_r})}+
 +b \sqrt{\barr{\gamma}} \, \extprod{\barr{D}^{\bar a} \barr{A}_r}{\barr{A}_r}
 %+\\
 %&
 %+b \sqrt{\barr{\gamma}}\, \barr{\gamma}^{\bar a \bar s}\, \barr{\gamma}^{\bar m \bar n}\, \extprod{\barr{A}_{\bar m}}{(\partial_{\bar n} \barr{A}_{\bar s}-\partial_{\bar s} \barr{A}_{\bar n}+\extprod{\barr{A}_{\bar n}}{\barr{A}_{\bar s}})}
 +\partial_{\bar m} (Y^{\bar m}\, \barr{\pi}^{\bar a})
 -\partial_{\bar m} Y^{\bar a} \, \barr{\pi}^{\bar m}
 +\extprod{\barr{\zeta}}{\barr{\pi}^{\bar a}} \,, %\nonumber
\end{align}
where angular indices are raised and lowered with the use of $\barr{\gamma}^{\bar a \bar b}$ and $\barr{\gamma}_{\bar a \bar b}$ respectively,
$\barr{F}_{\bar m \bar n} \eqdef 
\partial_{\bar m} \barr{A}_{\bar n}-\partial_{\bar n} \barr{A}_{\bar m}+\extprod{\barr{A}_{\bar m}}{\barr{A}_{\bar n}}$
and  $\barr{D}_{\bar a} \eqdef \barr{\nabla}_{\bar a} + \extprod{\barr{A}_{\bar a}}{}$ is the asymptotic gauge-covariant derivative, being $\barr{\nabla}_{\bar a}$ the covariant derivative on the unit round sphere.

One sees immediately that the asymptotic transformations above are affected only by the boost $b$ and the rotations $Y^{\bar m}$, but not by the translations $T$ and $W$.
Moreover, these transformations exhibit two main differences with respect to the analogous transformations in electrodynamics~\cite{Henneaux-ED}.
First, the radial and angular components of the fields do \emph{not} transform independently, due to the mixing terms in the transformation of the momenta.
Secondly, none of the asymptotic fields are gauge invariant.
Both these properties are a consequence of the non-abelian nature of the gauge group and will play an important role in the discussion of parity conditions in the next section.

\section{Well-defined Hamiltonian formulation and parity conditions} \label{sec:parity}

The fall-off conditions~(\ref{fall-off-conditions}) are not sufficient to ensure the finiteness of the symplectic form~(\ref{symplectic-form}), which is, indeed, still logarithmically divergent.
This problem can be fixed in the following way.
First, one assigns, independently to one another, a definite parity to the asymptotic part of the fields, $\barr{A}_r(\barr{x})$ and $\barr{A}_{\bar{a}}(\barr{x})$, so that they are either odd or even functions on the two-sphere.
Secondly, one imposes the opposite parity on the asymptotic part of the corresponding conjugated momenta, $\barr{\pi}^r(\barr{x})$ and $\barr{\pi}^{\bar a}(\barr{x})$.
This way, the logarithmically divergent term in the symplectic form is, in fact, zero once integrated on the two-sphere.

Specifically, let us assume that $\barr{A}_r$ has parity $s\in \mathbb{Z}_2$ and that $\barr{A}_{\bar{a}}$ has parity $\sigma \in \mathbb{Z}_2$, i.e., they behave under the antipodal map,\footnote{
In the usual angular coordinates $(\theta,\varphi)$, the antipodal map $\barr{x} \mapsto -\barr{x}$ corresponds explicitly to the transformation $(\theta,\varphi) \mapsto (\pi-\theta,\pi+\varphi)$. See~\cite{Henneaux-ED} for details.
}
denoted hereafter by $\barr{x} \mapsto -\barr{x}$, as 
\begin{equation} \label{parity-conditions-undetermined}
 \barr{A}_r (-\barr{x}) = (-1)^s \, \barr{A}_r (\barr{x})
 \qquad \text{and} \qquad
 \barr{A}_{\bar{a}} (-\barr{x}) = (-1)^\sigma \, \barr{A}_{\bar{a}} (\barr{x}) \,.
\end{equation}
Then, the symplectic form is made finite by assuming that $\barr{\pi}^r$ has parity $s+1$ and that $\barr{\pi}^{\bar a}$ has parity $\sigma+1$.
The key observation is that the values of $s$ and $\sigma$ are unequivocally determined by the requirement that the Poincar\'e transformations are canonical and that they preserve the parity transformations.
In electrodynamics, it is possible to relax the strict parity conditions leaving the symplectic form still finite~\cite{Henneaux-ED}.
We will review how this procedure works in electrodynamics in section~\ref{sec:relax-parity-ED} and attempt to apply it to the Yang-Mills case in section~\ref{sec:relax-parity-YM}.

\subsection{Proper and improper gauge transformations} \label{subsec:gauge-transformations}
Before we determine the parity conditions, let us extend the discussion of subsection~\ref{subsec:gauge-transformations-intro} and provide some more details about gauge transformations.
As we have already mentioned in subsection~\ref{subsec:gauge-transformations-intro}, gauge transformations are generated by
\begin{equation} \label{gauge-generator-again}
 G[\lambda] \eqdef \int d^3 x\, \sprod{\lambda (x)}{\mathscr{G} (x)} \,,
\end{equation}
which is functionally differentiable with respect to the canonical fields if, and only if, the surface term
\begin{equation} \label{gauge-generator-surface}
 \oint_{S^2_{\infty}} d^2 s_k \, \sprod{\lambda}{\pi^k}
 = \oint d^2 \barr{x} \; \sprod{\barr{\lambda}}{\barr{\pi}^r} 
\end{equation}
vanishes.
In the right-hand side of the above expression, we have inserted the fall-off behaviour of the fields~(\ref{fall-off-conditions}) and of the gauge parameter~(\ref{fall-off-conditions-gauge}).
Note that the integral on the right-hand side is an integral over a unit sphere, since the dependence on the radial coordinate $r$ disappears after taking the limit to an infinite-radius sphere in the left-hand side.
One sees immediately that the surface term vanishes for every allowed $\barr{\pi}^r$ if, and only if, the asymptotic gauge parameter $\barr{\lambda}$ has parity $s$, which is the opposite parity of $\barr{\pi}^r$.

There is an alternative way to make the generator~(\ref{gauge-generator-again}) differentiable.
Precisely, one defines the extended generator
\begin{equation} \label{gauge-generator-extended}
 G_{\text{ext.}}[\epsilon] \eqdef \int d^3 x \, \sprod{\epsilon(x)}{\mathscr{G}(x)}
 -\oint d^2 \barr{x}\; \sprod{\barr{\epsilon}(\barr{x})}{\barr{\pi}^r(\barr{x})} \,,
\end{equation}
where the function $\epsilon(x)$ is required to satisfy the same fall-off behaviour~(\ref{fall-off-conditions-gauge}) of $\lambda (x)$ and $\zeta(x)$, but its asymptotic part $\barr{\epsilon}$ is \emph{not} restricted to have a definite parity.
One can easily verify that $G_{\text{ext.}}[\epsilon]$ is functionally differentiable and that it generates the infinitesimal transformations
\begin{align} \label{gauge-transformations-improper1}
 \delta_\epsilon A_a &\eqdef \{ A_a, G_{\text{ext.}}[\epsilon] \} = -\partial_a \epsilon +\extprod{\epsilon}{A_a} \,, \\
 \label{gauge-transformations-improper2}
 \delta_\epsilon \pi^a &\eqdef \{ \pi^a, G_{\text{ext.}}[\epsilon] \} = \extprod{\epsilon}{\pi^a} \,.
\end{align}
Moreover, one can also verify that $\big\{ G_{\text{ext.}}[\epsilon], H \big\}=0$, so that $G_{\text{ext.}}[\epsilon]$ is the generator of a symmetry.
The infinitesimal transformations above can be integrated to get the transformations with parameter $\mathcal{U}\eqdef \exp (-\epsilon) \in \SU(N)$
\begin{align} \label{gauge-transformations-improper-full1}
 A_a \mapsto \Gamma_\mathcal{U} (A_a) & = \mathcal{U}^{-1} A_a \,\mathcal{U}+\mathcal{U}^{-1} \partial_a  \,\mathcal{U} \,, \\
 \label{gauge-transformations-improper-full2}
 \pi^a \mapsto \Gamma_\mathcal{U} (\pi^a)  &= \mathcal{U}^{-1} \pi^a \,\mathcal{U} \,,
\end{align}
where the products on the right-hand sides are products among matrices.

Note that, when $\barr{\epsilon}$ has parity $s$, the surface term in~(\ref{gauge-generator-extended}) vanishes and $G_{\text{ext.}}[\epsilon]$ coincides with $G[\epsilon]$
In this case, the symmetries generated by $G_{\text{ext.}}[\epsilon]$ are precisely the already-discussed gauge transformations connecting physically-equivalent field configurations.
We will refer to them in a rather pedantic way as \emph{proper gauge transformations}, in order to avoid any possible misunderstanding in the following discussion.

When $\barr{\epsilon}$ has parity $s+1$, the surface term in~(\ref{gauge-generator-extended}) does \emph{not} vanish any more.
The transformation generated by $G_{\text{ext.}}[\epsilon]$, in this case, connects physically-inequivalent field configurations.
We refer to this transformations as \emph{improper gauge transformations}, following~\cite{Teitelboim-YM2}.
These, on the contrary of proper gauge transformations, are true symmetry of the theory connecting physically-inequivalent field configurations.%\footnote{A detailed discussion about proper and improper gauge transformations is contained in~\cite{Teitelboim-YM2}}
A general transformation generated by $G_{\text{ext.}}[\epsilon]$ will be the combination of a proper gauge transformation and of an improper one.

The generator~(\ref{gauge-generator-extended}) is made of two pieces.
The former consists of the Gauss constraints $\mathscr{G}$ smeared with the function $\epsilon (x)$.
As a consequence, this term vanishes when the constraints are satisfied.
The latter is a surface term.
One can compute the value of the generator when the constraints are satisfied, which is, in particular, the case for any solution of the equations of motion.
One obtains
\begin{equation} \label{charge}
 G_{\text{ext.}}[\epsilon]
 \weq -\oint d^2 \barr{x}\; \sprod{\barr{\epsilon}(\barr{x})}{\barr{\pi}^r(\barr{x})}
 \defeq -Q[\barr{\epsilon}] \,,
\end{equation}
where we have defined the charge $Q[\barr{\epsilon}]$.
When the Lie-algebra-valued function $\epsilon(\barr{x})$ is constant over the sphere, we can write
$Q[\barr{\epsilon}] = \sprod{\barr{\epsilon}}{Q_0}$
in terms of the total colour charge measured at spatial infinity
\begin{equation} \label{colour-charge}
 Q_0 \eqdef \oint d^2 \barr{x} \, \pi^r (\barr{x}) \,.
\end{equation}

Finally, let us determine the transformation of the asymptotic fields under proper and improper gauge transformations.
Expanding the equations~(\ref{gauge-transformations-improper1}) and~(\ref{gauge-transformations-improper2}) using the fall-off conditions~(\ref{fall-off-conditions}), one finds
\begin{align} \label{gauge-transformations-improper-asymptotic}
 \delta_\epsilon \barr{A}_r = \extprod{\barr{\epsilon}}{\barr{A}_r} \,,
 &&
 \delta_\epsilon \barr{A}_{\bar a} = -\partial_{\bar a} \barr{\epsilon} +\extprod{\barr{\epsilon}}{\barr{A}_{\bar a}} \,,
 &&
 \delta_\epsilon \barr{\pi}^r = \extprod{\barr{\epsilon}}{\barr{\pi}^r} \,,
 &&
 \delta_\epsilon \barr{\pi}^{\bar a} = \extprod{\barr{\epsilon}}{\barr{\pi}^{\bar a}} \,,
\end{align}
whereas, expanding the equations~(\ref{gauge-transformations-improper-full1}) and~(\ref{gauge-transformations-improper-full2}), one finds
\begin{align} \label{gauge-transformations-improper-full-asymptotic1}
 \Gamma_{\mathcal{U}} (\barr{A}_r) &= \barr{\mathcal{U}}^{-1} \barr{A}_r \, \barr{\mathcal{U}} \,,
 &
 \Gamma_{\mathcal{U}} ( \barr{A}_{\bar a}) &=\barr{\mathcal{U}}^{-1} \barr{A}_{\bar a} \, \barr{\mathcal{U}} +\barr{\mathcal{U}}^{-1} \partial_{\bar a} \, \barr{\mathcal{U}} \,,
 \\ \label{gauge-transformations-improper-full-asymptotic2}
 \Gamma_{\mathcal{U}} ( \barr{\pi}^r) &= \barr{\mathcal{U}}^{-1} \barr{\pi}^r \, \barr{\mathcal{U}}\,,
 &
\Gamma_{\mathcal{U}} ( \barr{\pi}^{\bar a} ) &= \barr{\mathcal{U}}^{-1} \barr{\pi}^{\bar a} \, \barr{\mathcal{U}} \,,
\end{align}
where $\barr{\mathcal{U}}\eqdef \exp(-\barr{\epsilon})$.
Note that the total colour charge transforms non-trivially under proper and improper gauge transformations as
\begin{equation} \label{colour-charge-transformation}
 \Gamma_{\mathcal{U}} (Q_0) =
 \oint d^2 \barr{x} \; \barr{\mathcal{U}}^{-1} (\barr{x}) \,  \pi^r (\barr{x}) \, \barr{\mathcal{U}} (\barr{x}) \,.
\end{equation}
We will complete this discussion once that we have determined the parity conditions in the next subsection.
% We will use these equations to determine the parity conditions in the next subsection.

\subsection{Poincar\'e transformations and parity conditions} \label{subsec:poincare-parity}
In this subsection, we elaborate on some aspects of the Poincar\'e transformations, that were left aside in the previous discussions in section~\ref{sec:Lorentz-fall-off} and we determine the parity conditions of the asymptotic fields, that is the values of $s$ and $\sigma$, which were introduced at the beginning of this section.
In order to do so, we require the Poincar\'e transformations to be canonical and to preserve the parity conditions.

To this end, let us take into consideration the asymptotic Poincar\'e transformations (\ref{poincare-asymptotic1})--(\ref{poincare-asymptotic4}).
The parts of the transformations depending on $\barr{\zeta}$ are in fact a proper gauge transformation, which we will discuss below.
The rest of the transformations preserves parity conditions as long as $\sigma=1$, as one can easily check.
%This is, of course, compatible with the conditions inferred above from proper gauge transformations.

Let us now impose that the Poincar\'e transformations are canonical.
This is achieved by imposing that $ \liephase_X \Omega= 0 $ or, equivalently,
\begin{equation} \label{canonical-definition}
 \extder (i_X \Omega)=0 \,,
\end{equation}
where $\Omega$ is the symplectic form~(\ref{symplectic-form}), $\liephase$ is the Lie derivative in phase space, and $X$ is the vector field in phase space defining the Poincar\'e transformations.
The left-hand side of the above expression is
\[
 \extder (i_X \Omega)=
 \int d^3 x \Big[
 \sprod{\extder \big( \delta_{\xi,\zeta} \pi^a \big) \wedge}{\extder A_a}+
 \sprod{\extder  \pi^a  \wedge}{\extder \big( \delta_{\xi,\zeta} A_a \big) }
 \Big] \,.
\]
This expression can be evaluated by inserting the explicit value of the transformations~(\ref{poincare-transformations1}) and~(\ref{poincare-transformations2}), together with~(\ref{poincare-xi}).
After a few lines of calculations and after the use of the fall-off conditions~(\ref{fall-off-conditions}), one finds
\begin{equation} \label{poincare-canonical-final}
 \extder (i_X \Omega)= \oint d^2 \barr{x} \; b \, \sqrt{\barr{\gamma}} \,
 \sprod{\extder \barr{A}_{\bar m} \wedge}{\extder ( \barr{D}^{\bar m} \barr{A}_r )} \,.
\end{equation}
One can note three things.
First, after the fall-off conditions have been imposed, the only part of the Poincar\'e transformations which could lead to some problem is the boost sector.
Secondly, the above expression is precisely the non-abelian analogous of the one derived in~\cite{Henneaux-ED} for electrodynamics.
Lastly, if $\sigma = 1$, the right-hand side vanishes as long as $s=0$, which fully determines the parity conditions.

In short, the asymptotic fields need to satisfy the parity conditions
\begin{align} \label{parity-conditions}
 \barr{A}_r (-\barr{x}) = \barr{A}_r (\barr{x}) \,,
 &&
 \barr{A}_{\bar{a}} (-\barr{x}) = -\barr{A}_{\bar{a}} (\barr{x}) \,, &&
 \barr{\pi}^r (-\barr{x}) = - \barr{\pi}^r (\barr{x}) \,,
 &&
 \barr{\pi}^{\bar{a}} (-\barr{x}) =  \barr{\pi}^{\bar{a}} (\barr{x}) \,.
\end{align}
Moreover, the gauge parameter of proper gauge transformations satisfies
\begin{equation} \label{parity-condition-gauge}
 \barr{\epsilon}_{\text{proper}} (-\barr{x})=
 \barr{\epsilon}_{\text{proper}} (\barr{x}) \,.
\end{equation}
It is easy to check that proper gauge transformations --- including the parts of the Poincar\'e transformations~(\ref{poincare-asymptotic1})--(\ref{poincare-asymptotic4}) depending on $\barr{\zeta}$ term --- preserve the parity conditions.

The parity conditions~(\ref{parity-conditions}) and~(\ref{parity-condition-gauge}) on the fields have a few consequences, other than making the symplectic form~(\ref{symplectic-form}) finite.
First, the Hamiltonian~(\ref{Hamiltonian}) is finite and functionally differentiable, as one can easily check.
With the exclusion of term containing the Gauss constraint, this was already true after that we had imposed the fall-off conditions~(\ref{fall-off-conditions}).
The parity conditions make it true also for the last term.

Second, improper gauge transformations are, at this stage, \emph{not} allowed.
Indeed, they change the asymptotic fields as in~(\ref{gauge-transformations-improper-asymptotic}) when the asymptotic part of the gauge parameter has parity
\begin{equation}
 \barr{\epsilon}_{\text{improper}} (-\barr{x})=
 -\barr{\epsilon}_{\text{improper}} (\barr{x}) \,.
\end{equation}
However, these transformations do not preserve the parity conditions~(\ref{parity-conditions}).
Therefore, if they were allowed, they would transform one point of the space of allowed field configurations to a point that does not belong to this space any more, which is not possible.
In subsection~\ref{subsec:loosen-parity}, we will discuss whether or not it is possible to modify parity conditions in order to restore the improper gauge transformations into the theory.

Third, the Poincar\'e transformations are canonical.
Their generator, which is presented in the next subsection, is finite and functionally differentiable.
Note that, with the exception of the boost, the transformations were already canonical even before imposing the parity conditions.
The parity conditions presented in this section fix the behaviour of the boost.

Last but not least, since $\barr{\pi}^r (\barr{x})$ is an odd function of $\barr{x}$, all the charges $Q[\barr{\epsilon}]$ defined in~(\ref{charge}) are vanishing when $\barr{\epsilon} (\barr{x})$ is an even function.
Notably, this includes the total colour charge $Q_{0}$, defined in~(\ref{colour-charge}), which is therefore zero.
Note that, despite the colour charge is not a gauge-invariant quantity, the statement that it is actually equal to zero is a gauge-invariant statement.
Indeed, using equation~(\ref{colour-charge-transformation}), one sees that the colour charge vanishes for every gauge transformation
$\barr{\mathcal{U}} (\barr{x}) = \exp \big[-\epsilon_{\text{proper}} (\barr{x}) \big]$,
after imposing the parity conditions~(\ref{parity-conditions}) and~(\ref{parity-condition-gauge}).

The above considerations would suggest that there are some issues if one wants a well-defined Lorentz boost and a non-zero colour charge in the Yang-Mills theory. A similar suggestion, coming from a different approach, 
was already present in~\cite{Christodoulou.Murchadha:1981}, where Christodoulou 
and {\'o} Murchadha studied the boost problem in General Relativity and 
briefly commented, at the end of their section 5, that the boost problem
does not seem to have  solutions for charged configurations 
in the Yang-Mills case, quite in contrast to the behaviour of 
General Relativity.

\subsection{Poincar\'e generator and algebra}

Now that we know that the Poincar\'e transformations are canonical, we present their finite and functionally-differentiable canonical generator, included the needed boundary term.
This is obtained from~(\ref{diffeo-generator}), in the particular case in which $\xi^\perp$ and $\xi^i$ are the ones in~(\ref{poincare-xi}).
After having reassembled and renamed the various terms, one finds
\begin{equation} \label{poincare-generator}
 P[\xi^\perp,\xi^i,\zeta]=
 \int d^3 x \, \big[
 \xi^\perp \, \mathscr{P}_0
 + \xi^i \, \mathscr{P}_i
 + \sprod{\zeta}{\mathscr{G}}
 \big]
 +\oint d^2 \barr{x} \; \mathscr{B} \,,
\end{equation}
where the generator of the normal component of the Poincar\'e transformations is
\begin{equation} \label{P0}
 \mathscr{P}_0 = \frac{\sprod{\pi^a}{\pi_a}}{2\sqrt{g}} + \frac{\sqrt{g}}{4} \sprod{F_{ab}}{F^{ab}} \,,
\end{equation}
the generator of the tangential component is
\begin{equation} \label{Pi}
 \mathscr{P}_i = \sprod{\pi^a}{\partial_i A_a} - \partial_a (\sprod{\pi^a}{A_i}) \,,
\end{equation}
the generator of the proper gauge transformations $\mathscr{G}$ is the Gauss constraint~(\ref{gauss-constraint}),and the explicit expression of the boundary term is
\begin{equation} \label{poincare-boundary}
 \mathscr{B} = \sprod{\barr{\pi}^r}{Y^{\bar a} \barr{A}_{\bar a}} \,,
\end{equation}
which is needed to make the generator~(\ref{poincare-generator}) functionally differentiable with respect to the canonical fields.

Finally, the Poincar\'e generator satisfy the algebra
\begin{equation}
 \Big\{ P \big[\xi_1^\perp, \xi_1^i, \zeta_1 \big], P \big[\xi^\perp_2, \xi_2^i, \zeta_2 \big] \Big\}=
 P \big[\widehat \xi^\perp, \widehat \xi, \widehat \zeta \big]\,,
\end{equation}
where
\begin{align}
 %% Normal deformation
 \widehat \xi^\perp &= \xi^i_1 \partial_i \xi^\perp_2 -\xi^i_2 \partial_i \xi^\perp_1 \,, \\
 %% Tangential deformation
 \widehat \xi^i &= g^{ij} (\xi^\perp_1 \partial_j \xi^\perp_2-\xi^\perp_2 \partial_j \xi^\perp_1)
 + \xi^j_1 \partial_j \xi^i_2 - \xi^j_2 \partial_j \xi^i_1 \,, \\
 %% Gauge
 \widehat \zeta &= A_i g^{ij} (\xi^\perp_1 \partial_j \xi^\perp_2-\xi^\perp_2 \partial_j \xi^\perp_1)
 +\xi^i_1 \partial_i \zeta_2 - \xi^i_2 \partial_i \zeta_1
 + \extprod{\zeta_1}{\zeta_2} \,.
\end{align}

This concludes this section, in which we have shown that imposing the fall-off conditions~(\ref{fall-off-conditions}) together with the parity conditions~(\ref{parity-conditions}) lead to a well-defined symplectic form with a well-defined Hamiltonian and a well-defined canonical action of the Poincar\'e group on the fields.
Moreover, enforcing the parity conditions~(\ref{parity-conditions}) has two consequences other than the ones listed above.
First, the improper gauge transformations are not allowed any more and, as a result, the asymptotic symmetry group is trivial.
Secondly, some of the charges~(\ref{charge}) measured at spatial infinity, and in particular the $Q[\barr{\epsilon}]$ with even $\barr{\epsilon}$, are vanishing.
Notably, this includes the total colour charge $Q_0$.\footnote
{ \label{foot:radial-parity}
In order to have a non-vanishing colour charge, we would need the radial components to satisfy the opposite parity conditions to the ones presented in this section.
However, these would make the Poincar\'e transformations non canonical.
Whether or not there is a way to implement the different parity conditions leaving the Poincar\'e transformations canonical will be discussed in the next section.
In addition, these parity conditions would also exclude the possibility of making proper gauge transformations with a non-vanishing part at infinity, but would allow improper gauge transformations.
}
In the next section, we explore the possibility of modifying the parity conditions, in order to restore improper gauge transformations as symmetries of the theory.

\section{Relaxing parity conditions and asymptotic symmetries in electrodynamics} \label{sec:relax-parity-ED}

In the previous analysis, we have imposed fall-off and parity conditions on the canonical fields and we have obtained, as a result, a well-defined Hamiltonian picture. 
However, at least in the case of electrodynamics, it is possible to weaken the parity conditions so that the symplectic form is still finite and improper gauge transformations are allowed, as it was shown in~\cite{Henneaux-ED}.
Before we investigate this possibility in the case of Yang-Mills, let us briefly show, in this section, how the procedure works in the simpler case of electrodynamics.

\subsection{Relaxing parity conditions} \label{subsec:loosen-parity-ED}
To begin with, let us note that the equations of the electromagnetic case can be inferred from the equations of this paper by replacing formally the one-form Yang-Mills potential $A_a$ with one-form electromagnetic potential $A_a^{\text{ED}}$ and the Yang-Mills conjugated momentum $\pi^a$ with the electromagnetic conjugated momentum $\pi^a_{ED}$.
In addition, one also needs to replace the Killing scalar product $\sprod{}{}$ with the product among real numbers and set to zero every term containing the non-abelian contributions given by $\extprod{}{}$.
In the remainder of this section, we will not write explicitly the subscript and the superscript ``ED'' on the fields, since we will consider only the electromagnetic case.

If we followed the same line of argument of section~\ref{sec:Lorentz-fall-off} in the case of electrodynamics, we would arrive at the same fall-off conditions~(\ref{fall-off-conditions}) and~(\ref{fall-off-conditions-gauge}) for the canonical fields and the gauge parameter, respectively.
These are precisely the fall-off conditions presented in~\cite{Henneaux-ED}.

Then, if we determined the parity conditions with the same reasoning of the section~\ref{sec:parity}, we would find out that any choice of definite parity for $\barr{A}_r$ and $\barr{A}_{\bar a}$ would be preserved by the Poincar\'e transformations.
However, these would be canonical only if the parity of $\barr{A}_r$ were opposite to that of $\barr{A}_{\bar a}$.
At this point, we choose the parity of $\barr{\pi}^r$ to be even, so that Coulomb is an allowed solution.
Therefore, we arrive at the parity conditions
\begin{align} \label{parity-conditions-ED}
 \barr{A}_r (-\barr{x})&= -\barr{A}_r (\barr{x}) \,,
 &
 \barr{\pi}^r (-\barr{x})&= \barr{\pi}^r (\barr{x}) \,,
 &
 \barr{A}_{\bar{a}} (-\barr{x})&= \barr{A}_{\bar{a}} (\barr{x}) \,,
 &
 \barr{\pi}^{\bar{a}} (-\barr{x})&= -\barr{\pi}^{\bar{a}} (\barr{x}) \,.
\end{align}
One consequence of these parity conditions is that the improper gauge transformations are not allowed, since they would add an odd part to the even $\barr{A}_{\bar a}$.
However, this issue can be easily solved by requiring that the fields satisfy the parity conditions given above up to an improper gauge transformation.
That is, we ask the field to satisfy the slightly weaker parity conditions
\begin{align} \label{parity-conditions-loosen-ED}
 \barr{A}_r &= \barr{A}_r^{\text{odd}} \,,
 &
 \barr{\pi}^r &= \barr{\pi}^r_{\text{even}} \,,
 &
 \barr{A}_{\bar{a}} &= \barr{A}_{\bar{a}}^{\text{even}}
 -\partial_{\bar a} \barr{\Phi}^{\text{even}} \,,
 &
 \barr{\pi}^{\bar{a}} &=  \barr{\pi}^{\bar{a}}_{\text{odd}} \,,
\end{align}
where $\barr{\Phi}^{\text{even}}(\barr{x})$ is an even function on the sphere.
With these parity conditions, the symplectic form is not finite any more.
Indeed, it contains the logarithmically divergent contribution
\[
 \int \frac{dr}{r}\oint_{S^2} d^2 \barr{x} \; \extder \barr{\pi}^a \wedge \extder \barr{A}_a=
 -\int \frac{dr}{r}\oint_{S^2} d^2 \barr{x} \; \extder \barr{\pi}^{\bar a} \wedge \extder \partial_{\bar a} \barr{\Phi}^{\text{even}}=
 \int \frac{dr}{r}\oint_{S^2} d^2 \barr{x} \; \extder \partial_{\bar a} \barr{\pi}^{\bar a} \wedge \extder  \barr{\Phi}^{\text{even}} \,,
\]
where we have integrated by parts in the last passage.
As it was noted in~\cite{Henneaux-ED}, supplementing the parity conditions~(\ref{parity-conditions-loosen-ED}) with the further condition
\begin{equation} \label{asymptotic-Gauss-ED}
 \partial_{\bar a} \barr{\pi}^{\bar a}=0 \,,
\end{equation}
which is nothing else than the asymptotic part of the Gauss constraint, makes the symplectic form finite without excluding any potential solution of the equations of motion.

Furthermore, one notes that also the alternative parity conditions
\begin{align} \label{parity-conditions-loosen-ED-alternative}
 \barr{A}_r &= \barr{A}_r^{\text{odd}} \,,
 &
 \barr{\pi}^r &= \barr{\pi}^r_{\text{even}} \,,
 &
 \barr{A}_{\bar{a}} &= \barr{A}_{\bar{a}}^{\text{odd}}
 -\partial_{\bar a} \barr{\Phi}^{\text{odd}} \,,
 &
 \barr{\pi}^{\bar{a}} &=  \barr{\pi}^{\bar{a}}_{\text{even}} \,,
\end{align}
supplemented with~(\ref{asymptotic-Gauss-ED}) lead to a finite symplectic form while allowing improper gauge transformations.
Either the choice of~(\ref{parity-conditions-loosen-ED}) for the parity conditions or that of~(\ref{parity-conditions-loosen-ED-alternative}) supplemented with~(\ref{asymptotic-Gauss-ED}) provides a theory of electrodynamics, in which the symplectic form is finite and improper gauge transformations are allowed.
The former choice of parity conditions is preferable since the latter excludes the possibility of magnetic sources and leads generically to divergences in the magnetic field as one approaches future and past null infinity, as pointed out in~\cite{Henneaux-ED}.

% Finally, before we move to discuss the Yang-Mills case in the next subsection, let us mention that relaxing the parity conditions in electrodynamics causes the action of the Poincar\'e group, and in particular the Lorentz boosts, to be non-canonical.
% The solution to this problem and its implications are thoroughly discussed in~\cite{Henneaux-ED}.
% This fact, however, will not be relevant in the following analysis of the Yang-Mills case.

\subsection{Making Poincar\'e transformations canonical} \label{subsec:poincare-canonical-ED}
The extended parity conditions~(\ref{parity-conditions-loosen-ED}) and~(\ref{parity-conditions-loosen-ED-alternative}) come with the advantage of including improper gauge transformations as symmetries of the theory at the cost, however, of making the Poincar\'e transformations non canonical.
Indeed, with these relaxed parity conditions, the left-hand side of~(\ref{poincare-canonical-final}) does not vanish any more.
The solution to this issue, presented in full details by Henneaux and Troessaert in~\cite{Henneaux-ED}, works as follows.

One introduces a new scalar field $\Psi$ and its corresponding canonical momentum $\pi_\Psi$, which is a scalar density of weight one.
In radial-angular coordinates, the scalar field and its canonical momentum are required to satisfy the fall-off conditions
\begin{align} \label{fall-off-Psi}
 \Psi= \frac{1}{r} \barr{\Psi} (\barr{x}) +\bigo (1/r^2) &&
 \text{and} &&
 \pi_{\Psi}= \frac{1}{r} \pi^{(1)}_\Psi (\barr{x}) +\smallo (1/r) \,.
\end{align}
Note that one assumes that the subleading contributions of  scalar field $\Psi$ are $\bigo(1/r^2)$, i.e. vanishing as $r$ tends to infinity at least as fast as $1/r^2$.
At the same time, one assumes that the subleading contributions of the momentum $\pi_\Psi$ are only $\smallo(1/r)$, i.e. vanishing faster than $1/r$, but not necessarily as fast as $1/r^2$.
Moreover, one imposes the constraint
\begin{equation}
 \pi_\Psi \approx 0 \,,
\end{equation}
so that the scalar field $\Psi$ is pure gauge in the bulk.
At this point, one modifies the symplectic form to
\begin{equation}
 \Omega= \int d^3 x \, \big[ \extder \pi^a \wedge \extder A_a
 +\extder \pi_\Psi \wedge \extder \Psi \big] +\omega \,,
\end{equation}
which contains the standard contributions in the bulk and, in addition, the non-trivial surface term
\begin{equation}
 \omega \eqdef \oint d^2 \barr{x} \, \sqrt{\barr{\gamma}} \, \extder \barr{\Psi} \wedge \extder \barr{A}_r \,.
\end{equation}
Finally, one extends the Poincar\'e transformations to
\begin{align}
 \delta_{\xi,\zeta} A_a &
 =\xi^\perp \frac{\pi_a}{\sqrt{g}}+\xi^i \partial_i A_a+\partial_a \xi^i A_i+\partial_a ( \xi^\perp \Psi -\zeta) \,,	\\
 %%%%
 %\label{poincare-transformations2}
 \delta_{\xi,\zeta} \pi^a &=
   \partial_b(\sqrt{g}\, \xi^\perp F^{ba}) +\xi^\perp \nabla^a \pi_\Psi
 +\partial_i (\xi^i \pi^a) - \partial_i \xi^a \pi^i \,, \\
 %%%%
 \delta_{\xi,\zeta} \Psi &= \nabla^a (\xi^\perp A_a) + \xi^i \partial_i \Psi \,, \\
 %%%%
 \delta_{\xi,\zeta} \pi_\Psi &= \xi^\perp \partial_a \pi^a + \partial_i (\xi^i \pi_\Psi) \,.
\end{align}
Note that, up to gauge transformations and to constraints, the first two equations are the usual Poincar\'e transformations of $A_a$ and $\pi^a$.
It is now straightforward to show that the symplectic form is finite, that the fall-off conditions are preserved under Poincar\'e transformations, and that these latter are canonical.

% \textbf{CASE B1}

In this paper, we present also an alternative way to achieve the same result. 
First we introduce a one-form $\phi_a$ and the corresponding canonical momentum $\Pi^a$, which is a vector density of weight one.
In polar coordinates, these new fields are required to satisfy the fall-off conditions
\begin{align} \label{fall-off-phi}
 \phi_r &=  \barr{\phi}_r (\barr{x}) +\bigo(1/r) \,, &
 \phi_{\bar a} &= r \barr{\phi}_{\bar a} (\barr{x}) +\bigo(r^0) \,, \\
 \Pi^r &= \frac{1}{r^2} \Pi^r_{(1)} (\barr{x}) +\smallo(1/r^2) \,, &
 \Pi^{\bar a} &= \frac{1}{r^3} \Pi^{\bar a}_{(1)} (\barr{x}) +\smallo(1/r^3) \,.
\end{align}
Note, as before, the different requirements for the subleading contributions of the field ($\bigo$) and of the momentum ($\smallo$).
Furthermore, we also impose the constraints
\begin{equation}
 \Pi^a \approx 0 \,,
\end{equation}
so that the new field $\phi_a$ is pure gauge in the bulk, and we modify the symplectic form to
\begin{equation}
 \Omega'= \int d^3 x \, \big[ \extder \pi^a \wedge \extder A_a
 +\extder \Pi^a \wedge \extder \phi_a \big] +\omega' \,,
\end{equation} 
which contains the non-trivial surface term
\begin{equation}
 \omega' \eqdef \oint d^2 \barr{x} \, \sqrt{\barr{\gamma}}\, \extder  (2 \barr{\phi}_r + \barr{\nabla}^{\bar a} \barr{\phi}_{\bar a}) \wedge \extder \barr{A}_r \,.
\end{equation}
Finally, one extends the Poincar\'e transformations to
\begin{align}
 %%%% A_a
 \delta_{\xi,\zeta} A_a &
 =\xi^\perp \frac{\pi_a}{\sqrt{g}}+\xi^i \partial_i A_a+\partial_a \xi^i A_i+\partial_a ( \xi^\perp \nabla^i \phi_i -\zeta) \,,	\\
 %%%% \pi^a
 %\label{poincare-transformations2}
 \delta_{\xi,\zeta} \pi^a &=
   \partial_b(\sqrt{g}\, \xi^\perp F^{ba}) -\xi^\perp \Pi^a
 +\partial_i (\xi^i \pi^a) - \partial_i \xi^a \pi^i \,, \\
 %%%% \phi_a
 \delta_{\xi,\zeta} \phi_a &= \xi^\perp A_a + \xi^i \partial_i \phi_a+\partial_a \xi^i \phi_i \,, \\
 %%%% \Pi^a
 \delta_{\xi,\zeta} \Pi^a &= -\nabla^a (\xi^\perp \partial_i \pi^i) + \partial_i (\xi^i \Pi^a) - \partial_i \xi^a \Pi^i \,.
\end{align}
Again, note that, up to gauge transformations and to constraints, the first two equations are the usual Poincar\'e transformations of $A_a$ and $\pi^a$.
Moreover, the symplectic form is finite, the fall-off conditions are preserved under Poincar\'e transformations, and these latter are canonical.

\subsection{Asymptotic algebra}
In this subsection, we compute the asymptotic algebras of the two cases presented in the previous section and we show that these are equivalent.
% \textbf{CASE A}

The first case, which introduces the scalar field $\Psi$ and its momentum $\pi_\Psi$, is the solution presented in~\cite{Henneaux-ED}.
The Poincar\'e transformations are generated by
\begin{equation} \label{poincare-generator-em1}
 P^{(1)}[\xi^\perp,\xi^i]=
 \int d^3 x \, \big[
 \xi^\perp \, \mathscr{P}_0^{(1)}
 + \xi^i \, \mathscr{P}_i^{(1)}
 \big]
 +\oint d^2 \barr{x} \; \mathscr{B}^{(1)} \,,
\end{equation}
where the generator of the normal component is
\begin{equation} \label{P0-em1}
 \mathscr{P}_0^{(1)} = \frac{\pi^a \pi_a}{2\sqrt{g}} + \frac{\sqrt{g}}{4} F_{ab}F^{ab}
 -\Psi \partial_a \pi^a -A_a \nabla^a \pi_\Psi \,,
\end{equation}
the generator of the tangential component is
\begin{equation} \label{Pi-em1}
 \mathscr{P}_i^{(1)} = \pi^a\partial_i A_a - \partial_a (\pi^a A_i) +\pi_\Psi \partial_i \Psi \,,
\end{equation}
the generator of the proper gauge transformations  is the Gauss constraint $\mathscr{G} = \partial_a \pi^a$,and the explicit expression of the boundary term is
\begin{equation} \label{poincare-boundary-em1}
 \mathscr{B}^{(1)} = b \left( \barr{\Psi} \barr{\pi}^r + \sqrt{\barr{\gamma}} \, \barr{A}_{\bar a} \barr{\nabla}^{\bar a} \barr{A}_r \right)
 +Y^{\bar a} \left( \barr{\pi}^r \barr{A}_{\bar a} + \sqrt{\barr{\gamma}}\, \barr{\Psi} \partial_{\bar a} \barr{A}_r \right)
 \,,
\end{equation}
which is needed to make the generator~(\ref{poincare-generator-em1}) functionally differentiable with respect to the canonical fields.
In addition, the proper and improper gauge symmetries are generated by
\begin{equation}
 G_{\epsilon,\mu}^{(1)} = \int d^3 x\, \big( \epsilon \, \mathscr{G}
 + \mu \, \pi_\Psi \big)
 -\oint d^2 \barr{x} \,\big( \barr{\epsilon} \, \barr{\pi}^r +\sqrt{\barr{\gamma}} \, \barr{\mu} \, \barr{A}_r \big) \,,
\end{equation}
which, together with~(\ref{poincare-generator-em1}), satisfies the algebra
\begin{align} \label{asymptotic-algebra-ED1}
 \big\{ P_{\xi_1^\perp,\xi_1}^{(1)} , P_{\xi_2^\perp,\xi_2}^{(1)} \big\} = P^{(1)}_{\widehat \xi^\perp, \widehat \xi} \,,
 &&
 \big\{ G_{\epsilon,\mu}^{(1)}, P_{\xi^\perp,\xi}^{(1)} \big\}= G_{\widehat \epsilon, \widehat \mu}^{(1)} \,,
 &&
 \big \{ G_{\epsilon_1,\mu_1}^{(1)}, G_{\epsilon_2,\mu_2}^{(1)} \big\}=0 \,,
\end{align}
where
\begin{align}
 \widehat \xi^\perp &=\xi^i_1 \partial_i \xi_2^\perp-\xi^i_2 \partial_i \xi_1^\perp \,,
 &
 \widehat \xi^i &= g^{ij} (\xi^\perp_1 \partial_j \xi^\perp_2-\xi^\perp_2 \partial_j \xi^\perp_1)
 + \xi^j_1 \partial_j \xi^i_2 - \xi^j_2 \partial_j \xi^i_1 \,, \\
 \widehat \mu &= \nabla^i (\xi^\perp \partial_i \epsilon) -\xi^i \partial_i \mu \,, &
 \widehat \epsilon &= \xi^\perp \mu - \xi^i \partial_i \epsilon \,.
\end{align}

% \textbf{CASE B1}
In the second case presented in the previous subsection, which introduces the one-form $\phi_a$ and its momentum $\Pi^a$, the Poincar\'e transformations are generated by
\begin{equation} \label{poincare-generator-em2}
 P^{(2)}[\xi^\perp,\xi^i]=
 \int d^3 x \, \big[
 \xi^\perp \, \mathscr{P}_0^{(2)}
 + \xi^i \, \mathscr{P}_i^{(2)}
 \big]
 +\oint d^2 \barr{x} \; \mathscr{B}^{(2)} \,,
\end{equation}
where the generator of the normal component is
\begin{equation} \label{P0-em2}
 \mathscr{P}_0^{(2)} = \frac{\pi^a \pi_a}{2\sqrt{g}} + \frac{\sqrt{g}}{4} F_{ab}F^{ab}
 -\nabla^a \phi_a \partial_b \pi^b+\Pi^a A_a \,,
\end{equation}
the generator of the tangential component is
\begin{equation} \label{Pi-em2}
 \mathscr{P}_i^{(2)} = \pi^a\partial_i A_a - \partial_a (\pi^a A_i) +
 \Pi^a\partial_i \phi_a - \partial_a (\Pi^a \phi_i) \,,
 %%%% alternative
 % \Pi_i \nabla^m \phi_m\,,
\end{equation}
the generator of the proper gauge transformations  is the Gauss constraint $\mathscr{G} = \partial_a \pi^a$, and the explicit expression of the boundary term is
\begin{equation} \label{poincare-boundary-em2}
 \mathscr{B}^{(2)} = b \big[(2 \barr{\phi}_r + \barr{\nabla}^{\bar a} \barr{\phi}_{\bar a}) \barr{\pi}^r + \sqrt{\barr{\gamma}} \, \barr{A}_{\bar a} \barr{\nabla}^{\bar a} \barr{A}_r \big]
 +Y^{\bar a} \, \barr{\pi}^r\, \barr{A}_{\bar a}
 \,,
\end{equation}
which is needed to make the generator~(\ref{poincare-generator-em2}) functionally differentiable with respect to the canonical fields.
In addition, the proper and improper gauge symmetries are generated by
\begin{equation}
 G_{\epsilon,\chi}^{(2)} = \int d^3 x\, \big( \epsilon \, \mathscr{G}
 + \chi_a \, \Pi^a \big)
 -\oint d^2 \barr{x} \,\big[ \barr{\epsilon} \, \barr{\pi}^r +\sqrt{\barr{\gamma}} \, (2 \barr{\chi}_r +\barr{\nabla}^{\bar a} \barr{\chi}_{\bar a}) \, \barr{A}_r \big] \,,
\end{equation}
which can be combined with~(\ref{poincare-generator-em2}) into the generator
\begin{equation}
 A^{(2)}[\xi^\perp,\xi,\epsilon,\chi_a] \eqdef
 P^{(2)}[\xi^\perp,\xi]+G^{(2)}[\epsilon,\chi] \,,
\end{equation}
satisfying the algebra
\begin{align} \label{asymptotic-algebra-ED2}
 \big\{
 A^{(2)}[\xi_1^\perp,\xi_1,\epsilon_1,\chi_1] ,
 A^{(2)}[\xi_2^\perp,\xi_2,\epsilon_2,\chi_2]
 \big\} =
 A^{(2)}[\hat{\xi}^\perp,\hat{\xi},\hat{\epsilon},\hat{\chi}] \,,
\end{align}
where
\begin{align}
 \widehat \xi^\perp =& \xi^i_1 \partial_i \xi_2^\perp-\xi^i_2 \partial_i \xi_1^\perp \,,
 \\
 \widehat \xi^i =& \tilde \xi^i + \xi^j_1 \partial_j \xi^i_2 - \xi^j_2 \partial_j \xi^i_1 \,,
 \\
 \tilde \xi^i \eqdef& g^{ij} (\xi^\perp_1 \partial_j \xi^\perp_2-\xi^\perp_2 \partial_j \xi^\perp_1) \,,
 \\
 \widehat \chi_a =& \xi_1^\perp \partial_a \epsilon_2-\xi_2^\perp \partial_a \epsilon_1
 + \xi_1^i \partial_i \chi^2_a - \xi_2^i \partial_i \chi^1_a
 +\tilde{\xi}_a \nabla^m \phi_m - \tilde{\xi}^m \partial_m \phi_a -\partial_a (\tilde{\xi}^m\phi_m)\,,
 \\
 \widehat \epsilon =& \xi_2^i \partial_i \epsilon_1 -\xi_1^i \partial_i \epsilon_2+
 \xi_2^\perp \nabla^a \chi^1_a -\xi_1^\perp \nabla^a \chi^2_a \,.
\end{align}

The asymptotic algebras~(\ref{asymptotic-algebra-ED1}) and~(\ref{asymptotic-algebra-ED2}) are equivalent.
To see this, one has to consider in the two cases the group of all the allowed transformations and take the quotient of it with respect to the proper gauge.
Only then, one can compare the brackets~(\ref{asymptotic-algebra-ED1}) and~(\ref{asymptotic-algebra-ED2}).
In the first case presented above, the proper gauge amount to those transformations for which $\barr{\epsilon}$ is odd and $\barr{\mu}$ is even.
In the second case presented above, the proper gauge amount to those transformations for which $\barr{\epsilon}$ is odd and
$\barr{\nabla \cdot \chi} \eqdef 2 \barr{\chi}_r + \barr{\nabla}^{\bar a} \barr{\chi}_{\bar a}$ is even.
The equivalence is then shown by identifying $\barr{\mu}$ with $\barr{\nabla \cdot \chi}$.

\section{Relaxing parity conditions and asymptotic symmetries in Yang-Mills} \label{sec:relax-parity-YM}
In this section, we try to apply the methods of the previous section to the non-abelian Yang-Mills case.
The goal is to obtain a Hamiltonian formulation of Yang-Mills with canonical Poincar\'e transformations and with allowed improper gauge transformations.
As we shall see, this goal cannot be entirely fulfilled.

\subsection{Relaxing parity conditions in Yang-Mills} \label{subsec:loosen-parity}

Let us now study the possibility of relaxing the parity conditions in Yang-Mills, in order to restore the improper gauge transformations also in this case.
Following the same line of argument of the electromagnetic case, we begin by requiring the asymptotic fields to satisfy the parity conditions~(\ref{parity-conditions}) up to asymptotic improper gauge transformations~(\ref{gauge-transformations-improper-full-asymptotic1}) and~(\ref{gauge-transformations-improper-full-asymptotic2}), so that
\begin{align} \label{parity-conditions-loosen1}
 \barr{A}_r &= \barr{\mathcal{U}}^{-1} \barr{A}_r^{\text{even}} \, \barr{\mathcal{U}}  \,,
 &
 \barr{\pi}^r &=  \barr{\mathcal{U}}^{-1} \barr{\pi}^r_{\text{odd}} \, \barr{\mathcal{U}} \,,
 \\ \label{parity-conditions-loosen2}
 \barr{A}_{\bar{a}} &= \barr{\mathcal{U}}^{-1} \barr{A}_{\bar{a}}^{\text{odd}} \barr{\mathcal{U}}
 + \barr{\mathcal{U}}^{-1} \partial_{\bar a}\, \barr{\mathcal{U}}   \,,
 &
 \barr{\pi}^{\bar{a}} &= \barr{\mathcal{U}}^{-1} \barr{\pi}^{\bar{a}}_{\text{even}} \, \barr{\mathcal{U}} \,,
\end{align}
where $\barr{\mathcal{U}} (\barr{x})=\exp \big[-\barr{\Phi}^{\text{odd}} (\barr{x}) \big] \in \SU(N)$ and the Lie-algebra-valued function $\barr{\Phi}^{\text{odd}} (\barr{x})$ is odd under the antipodal map $\barr{x} \mapsto -\barr{x}$.
Therefore, the Lie-group-valued function $\barr{\mathcal{U}} (\barr{x})$ behaves as $\barr{\mathcal{U}} (-\barr{x})= \barr{\mathcal{U}} (\barr{x})^{-1}$ under the antipodal map.
These new parity conditions introduce the logarithmically divergent part
\begin{equation} 
\label{log-div1}
\begin{split}
 &\int \frac{dr}{r} \oint_{S^2} d^2 \barr{x} \;
 \sprod{\extder \barr{\pi}^a \wedge}{ \extder \barr{A}_a}
 =\\
 =&\int \frac{dr}{r} \oint_{S^2} d^2 \barr{x} \;
 \Bigg\{
 \sprod{ \Big(\extder \barr{\mathcal{U}} \, \barr{\mathcal{U}}^{-1}   \Big)\wedge}
 {
 \extder \Big(
 \partial_{\bar a} \barr{\pi}^{\bar a}_{\text{even}}
 +\extprod{\barr{A}_r^{\text{even}}}{\barr{\pi}^r_{\text{odd}}}
 +\extprod{\barr{A}_{\bar a}^{\text{odd}}}{\barr{\pi}^{\bar a}_{\text{even}}}
 \Big)
 }
 +\\
 &
 -\frac{1}{2}
 \sprod{
 \Big[ \extprod{\Big(\extder \barr{\mathcal{U}} \, \barr{\mathcal{U}}^{-1}   \Big) \wedge}{\Big(\extder \barr{\mathcal{U}} \, \barr{\mathcal{U}}^{-1}   \Big)} \Big]
 }{
 \Big(
 \partial_{\bar a} \barr{\pi}^{\bar a}_{\text{even}}
 +\extprod{\barr{A}_r^{\text{even}}}{\barr{\pi}^r_{\text{odd}}}
 +\extprod{\barr{A}_{\bar a}^{\text{odd}}}{\barr{\pi}^{\bar a}_{\text{even}}}
 \Big)
 }
 \Bigg\}
\end{split}
\end{equation}
in the symplectic form,
%where we have already eliminated all the terms vanishing due to parity, performed an integration by parts, and simplified the expression.
%The precise derivation of~(\ref{log-div1})
whose precise derivation is presented in appendix~\ref{app:symplectic-form}.

At this point, we note that the second factor in both summands of the right-hand side of~(\ref{log-div1}) is nothing else than the asymptotic Gauss constraint $\barr{\mathscr{G}}_0$ evaluated when $\barr{\Phi}^{\text{odd}}=0$, which is related to the asymptotic Gauss constrain $\barr{\mathscr{G}}$ with non-vanishing $\barr{\Phi}^{\text{odd}}$ by the expression $\barr{\mathscr{G}} = \barr{\mathcal{U}}^{-1} \, \barr{\mathscr{G}}_0 \, \barr{\mathcal{U}}$, so that the one vanishes if, and only if, the other does.
Therefore, we can keep the symplectic form finite by restricting the phase space to those field configurations that satisfy, together with the fall-off conditions~(\ref{fall-off-conditions}) and the parity conditions~(\ref{parity-conditions-loosen1}) and~(\ref{parity-conditions-loosen2}), also the asymptotic Gauss constraint
\begin{equation} \label{asymptotic-Gauss-YM}
 \partial_{\bar a} \barr{\pi}^{\bar a}
 +\extprod{\barr{A}_r}{\barr{\pi}^r}
 +\extprod{\barr{A}_{\bar a}}{\barr{\pi}^{\bar a}}
=0 \,.
\end{equation}
Note that imposing this further condition does not exclude any of the former solutions to the equations of motion, since every solution was already satisfying the (asymptotic part of the) Gauss constraint.
This shows that it is possible to relax the parity conditions in order to allow improper gauge transformations, but nevertheless leaving the symplectic form finite.

In electrodynamics, one notes that it is possible to start with a different set of parity conditions and to relax them, so that the symplectic form is nevertheless finite.
These freedom, was used in section~\ref{subsec:loosen-parity-ED} in order to present two possibility for the parity of the angular components of the asymptotic part of the fields.\footnote
{
In principle, one could use the same freedom for the parity of the radial component of the asymptotic fields, but this was already fixed by the physical requirement that Coulomb is a solution.
}
One could wonder whether or not this freedom is present also in the Yang-Mills case.

First, one notes that picking the opposite parity for the angular part is problematic.
Specifically, the asymptotic part of the Poincar\'e transformations~(\ref{poincare-asymptotic1})--(\ref{poincare-asymptotic4}) contains the term $\barr{F}_{\bar a \bar b}$ and the operator
$\barr{D}_{\bar a} \eqdef \barr{\nabla}_{\bar a} + \extprod{\barr{A}_{\bar a}}{}$.
If we took $\barr{A}_{\bar a}$ to be of even parity (up to asymptotic proper/improper gauge transformations) we would end up with terms of indefinite parity after applying the Poincar\'e transformations.

Secondly, one could try to pick the opposite parity conditions for the radial components of the asymptotic fields (up to asymptotic proper/improper gauge transformations).
This choice would have the advantage of allowing a non-vanishing value of the colour charge, as discussed in footnote~\ref{foot:radial-parity}. 
However, for this choice, the method used above to make the symplectic form finite does not work any more even after imposing the asymptotic Gauss constraint.\footnote{ 
The method used to make the symplectic form finite in this subsection works if $\barr{A}_r$ and $\barr{A}_{\bar a}$ are chosen so that they have opposite parity when $\barr{\Phi}^{\text{odd}} = 0$.  Therefore, the method presented in this section would still work if we chose, at the same time, the opposite parity conditions both for the radial and for the angular components, with respect to those presented in~(\ref{parity-conditions-loosen1})--(\ref{parity-conditions-loosen2}).
However, we have already discussed that changing the parity conditions of the angular components leads to other issues.
}

To sum up, we have found a way of relaxing the strict parity conditions of section~\ref{sec:parity} in order to allow improper gauge transformations, but leaving the symplectic form finite.
We have also discussed why different choices for the parity conditions are less appealing and more problematic in Yang-Mills compared to electrodynamics.
As expected, the price to pay when relaxing the parity conditions is that the Poincar\'e transformations are not canonical any more.
We will discuss what can be done to fix this issue in the next subsection.

\subsection{Attempt to make the Poincar\'e transformations canonical} \label{subsec:attempt-Poincare-canonical}

In order to make the Poincar\'e transformations canonical the following expression, which is the Lie derivative of the symplectic form, has to vanish:
\begin{equation} 
\label{poincare-not-canonical}
 \liephase_X \Omega =
 \extder (i_X \Omega)=
 \oint_{S^2} d^2 \barr{x} \; 
b \, \sqrt{\barr{\gamma}} \, 
\barr{\gamma}^{\bar m \bar n} \,
\sprod{\extder \barr{A}_{\bar m} \wedge}{\extder ( 
\barr{D}^{\bar m} \barr{A}_r )} \,,
\end{equation}
possibly adding a surface term to the symplectic form and introducing new fields, which are non-trivial only at the boundary.
One could try  to follow the line of reasoning of section~\ref{subsec:poincare-canonical-ED} also in Yang-Mills.
Since the Lie derivative of the symplectic form fails to vanish due to the Lorentz boost, we will focus on the Lorentz boost and neglect the rest of the Poincar\'e transformations in the following.
In other words, we will consider the case in which $\xi^\perp = rb$ and $\xi^i = 0$.
Moreover, we discuss, separately, the possible implementation of each one of the two solutions presented in section~\ref{subsec:poincare-canonical-ED} and adapted to the Yang-Mills case.

\subsubsection{Case 1} \label{subsubsec:ansatz-Psi}
First, let us consider the solution described in section~\ref{subsec:poincare-canonical-ED} which uses the scalar field $\Psi$ and its conjugated momentum $\pi_\Psi$, first found in~\cite{Henneaux-ED}.
Also in the Yang-Mills case, we supplement the field with the fall-off conditions~(\ref{fall-off-Psi}), the further constraint $\pi_\Psi \approx 0$, and the symplectic structure in the bulk
\begin{equation}
 \Omega= \int d^3 x \, \big[
 \sprod{\extder \pi^a \, \wedge }{ \extder A_a}
 +\sprod{ \extder \pi_\Psi \, \wedge }{ \extder \Psi }
 \big] \,.
\end{equation}
Moreover, we impose the action of the Lorentz boost on the fields to be
\begin{align}
 \label{boost-Psi-in}
 \delta_{\xi^\perp} A_a &
 =\xi^\perp \frac{\pi_a}{\sqrt{g}}+ D_a ( \xi^\perp \Psi ) \,,	\\
 %%%%
 \delta_{\xi^\perp} \pi^a &=
   \partial_b(\sqrt{g}\, \xi^\perp F^{ba}) +\xi^\perp \nabla^a \pi_\Psi - \xi^\perp \extprod{\Psi}{\pi^a} \,, \\
 %%%%
 \delta_{\xi^\perp} \Psi &= \nabla^a (\xi^\perp A_a) \,, \\
 %%%%
 \label{boost-Psi-end}
 \delta_{\xi^\perp} \pi_\Psi &= \xi^\perp \mathscr{G} \,,
\end{align}
which preserve both the fall-off conditions and the constraints.
The above transformations \emph{would be} generated by
\begin{equation}
 P [\xi^\perp] \eqdef
 \int d^3 x \, \xi^\perp \left[
 \frac{\sprod{\pi^a}{\pi_a}}{2\sqrt{g}} + \frac{\sqrt{g}}{4} \sprod{F_{ab}}{F^{ab}}
 - \sprod{\Psi}{\mathscr{G}} - \sprod{A_a}{\nabla^a \pi_\Psi}
 \right]
 + (\text{boundary}) \,,
\end{equation}
\emph{if} a suitable boundary term existed, so that the generator above were functionally differentiable with respect to the canonical fields (as we shall see in the following, such boundary term does not exist).
Let us now denote with $X'$ the vector field in phase space defining the Lorentz boost~(\ref{boost-Psi-in})--(\ref{boost-Psi-end}) and let us define
\begin{equation}
 \omega_0 \eqdef
 \oint_{S^2} d^2 \barr{x} \, \sqrt{\barr{\gamma}} \, \sprod{\extder \barr{\Psi} \, \wedge}{ \extder \barr{A}_r} \,,
\end{equation}
such that one finds
\begin{equation} \label{lie-Omega-omega0}
 \liephase_{X'} (\Omega + \omega_0) =
 \oint_{S^2} d^2 \barr{x} \, b \sqrt{\barr{\gamma}} \,
 \left[
 \sprod{\extder \barr{A}_{\bar m} \,\wedge}{ \extder( \extprod{ \barr{A}^{\bar m} }{\barr{A}_r} ) } -
 \sprod{\extder \barr{\Psi} \,\wedge}{ \extder( \extprod{ \barr{\Psi} }{\barr{A}_r} ) }
 \right]
\end{equation}
At this point, one needs to find a second boundary term $\omega_1$, whose phase-space Lie derivative $\liephase_{X'} \omega_1$ is the opposite of the expression above.
However, one immediately faces the issue that even the first term inside square brackets of the expression above cannot be compensated by some expression contained in $\liephase_{X'} \omega_1$, for any $\omega_1$ built from the canonical fields.
Indeed, the first term in~(\ref{lie-Omega-omega0}) contains only the asymptotic part of the field $A$, without any derivative, but the asymptotic transformations of the fields under Lorentz boosts do not contain any such term.
In other words, one \emph{cannot} find an extra surface term to the symplectic structure $\omega \eqdef \omega_0 + \omega_1$, which is build from the given fields and satisfies $\liephase_{X'} (\Omega + \omega)=0$.

\subsubsection{Case 2} \label{subsubsec:ansatz}
Secondly, one could try to adapt to the Yang-Mills case the other solution described in section~\ref{subsec:poincare-canonical-ED}, namely the one introducing the one form $\phi_a$ and its conjugated momentum $\Pi^a$.
Also in this case, we supplement the fields with the fall-off conditions~(\ref{fall-off-phi}), the further constraints $\Pi^a \approx 0$, and the symplectic form in the bulk
\begin{equation} \label{Omega-bulk-phi}
 \Omega' = \int d^3 x \, \big[
 \sprod{ \extder \pi^a \, \wedge }{ \extder A_a}
 +\sprod{ \extder \Pi^a \, \wedge }{ \extder \phi_a }
 \big] \,.
\end{equation} 
Moreover, we impose the action of the Lorentz boost on the fields to be
\begin{align}
 %%%% A_a
 \label{boost-YM-phi-in}
 \delta_{\xi^\perp} A_a &
 =\xi^\perp \frac{\pi_a}{\sqrt{g}}+ D_a ( \xi^\perp \mathscr{D}^i \phi_i) \,,	\\
 %%%% \pi^a
 \delta_{\xi^\perp} \pi^a &=
   \partial_b(\sqrt{g}\, \xi^\perp F^{ba}) -\xi^\perp \Pi^a
   + \xi^\perp \extprod{ \pi^a }{ \mathscr{D}^i \phi_i }
   + \xi^\perp \, c \, \extprod{ \phi^a }{ \mathscr{G} } \,, \\
 %%%% \phi_a
 \delta_{\xi^\perp} \phi_a &= \xi^\perp A_a  \,, \\
 %%%% \Pi^a
 \label{boost-YM-phi-end}
 \delta_{\xi^\perp} \Pi^a &= -\mathscr{D}^a (\xi^\perp \mathscr{G})  \,,
\end{align}
where $\mathscr{D}_a \eqdef \nabla_a +c_1 \, \extprod{A_a}{}+ c_2 \, \extprod{\pi_a}{}$ and $c_1,c_2 \in \real$ are free parameters that one can set later to suitable values in order to make the Lorentz boost canonical.
One can verify that the above transformations preserve both the fall-off conditions and the constraints.
Moreover, they \emph{would be} generated by
\begin{equation} \label{boost-generator-phi}
 P' [\xi^\perp] \eqdef
 \int d^3 x \, \xi^\perp \left[
 \frac{\sprod{\pi^a}{\pi_a}}{2\sqrt{g}} + \frac{\sqrt{g}}{4} \sprod{F_{ab}}{F^{ab}}
 - \sprod{\mathscr{D}^a \phi_a }{\mathscr{G}} + \sprod{A_a}{\Pi^a}
 \right]
 + (\text{boundary}) \,,
\end{equation}
\emph{if} a suitable boundary term existed, so that the generator above were functionally differentiable with respect to the canonical fields (as we shall see in the following, such boundary term does not exist).
One can easily compute that
\begin{equation} \label{lie-symplectic-bulk}
 \liephase_{X'} \Omega' = 
 \oint_{S^2} d^2 \barr{x} \, b  
 \Big[
 \sqrt{\barr{\gamma}} \, \sprod{\extder \barr{A}_{\bar m} \,\wedge}{ \extder \big( D^{\bar m} \barr{A}_r \big) }
 + \sprod{ \extder \barr{\pi}^r \, \wedge}{ \extder \barr{\mathscr{D} \phi }}
 \Big]  \,,
\end{equation}
where
\[
\barr{\mathscr{D} \phi}\eqdef 2 \barr{\phi}_r + \barr{\nabla}^{\bar m} \barr{\phi}_{\bar m}
+ c_1 \, \big( \extprod{\barr{A}_r}{\barr{\phi}_r} + \extprod{ \barr{A}^{\bar m} }{ \barr{\phi}_{\bar m} } \big)
+ c_2 \, \big( \extprod{\barr{\pi}^r}{\barr{\phi}_r} + \extprod{ \barr{\pi}^{\bar m} }{ \barr{\phi}_{\bar m} } \big) 
\]
is the leading contribution in the expansion of $\mathscr{D}^a \phi_a =  \barr{\mathscr{D}\phi} /r + \bigo(1/r^2)$
and $X'$ is the vector field on phase space that defines the Lorentz boost~(\ref{boost-YM-phi-in})--(\ref{boost-YM-phi-end}).

One hopes that, with respect to the previous case concerning $\Psi$ and $\pi_\Psi$, one can now tackle the problem more efficiently, since there are now fields transforming asymptotically as the asymptotic part of $A_a$ without derivatives.
Namely, the one form $\phi_a$ transforms asymptotically under Lorentz boosts like
$\delta_{\xi^\perp} \barr{\phi}_{a} = b \, \barr{A}_a$.

In order to compensate for the terms contained in~(\ref{lie-symplectic-bulk}), we use the following ansatz for the boundary term of the symplectic form:
\begin{equation} \label{ansatz-omega}
 \begin{aligned}
 \omega' = \oint_{S^2} d^2 \barr{x} \; \sqrt{\barr{\gamma}} \,
 \Big[
 & a_0 \, \sprod{\extder \big( \barr{\nabla}^{\bar m}  \barr{\phi}_{\bar m} \big) \wedge}{\extder \barr{A}_r } 
 + a_1 \, \sprod{\extder \barr{\phi}_{r} \wedge}{\extder \barr{A}_r } 
 + a_2 \, \sprod{ \barr{A}_{r} }{ \extprod{ \extder \barr{\phi}^{\bar m} \, \wedge}{ \extder \barr{A}_{\bar m} } } + \\
 +& a_3 \, \sprod{ \barr{\phi}_{r} }{ \extprod{ \extder \barr{A}^{\bar m} \, \wedge}{ \extder \barr{A}_{\bar m} } } 
 + a_4 \, \sprod{ \barr{A}_{\bar m} }{ \extprod{ \extder \barr{A}^{\bar m} \, \wedge}{ \extder \barr{\phi}_{r} } } + \\
 +& a_5 \, \sprod{ \barr{A}_{\bar m} }{ \extprod{ \extder \barr{\phi}^{\bar m} \, \wedge}{ \extder \barr{A}_{r} } } 
 + a_6 \, \sprod{ \barr{\phi}_{\bar m} }{ \extprod{ \extder \barr{A}^{\bar m} \, \wedge}{ \extder \barr{A}_{r} } } 
 \Big] \,,
\end{aligned}
\end{equation}
where $a_0, \dots, a_6 \in \real$ are free parameters that can be set to a suitable value in order to achieve $\liephase_{X'} (\Omega' + \omega') = 0$.
Note that one has to restrict the possible values of the parameters $a_0, \dots, a_6$, in order to ensure that the two-form $\omega'$ is closed.
In any case, one can show that no value of the parameters $a_0, \dots, a_6$, $c_1$, and $c_2$ can be found in order to make the Lorentz boost canonical.
A more detailed discussion about the reasons why we used the ansatz above and the computations needed to show that no value of the free parameters make the Lorentz boost canonical can be found in appendix~\ref{app:details-computations}.

In conclusion, we were not able to find a solution to the problem of making the Poincar\'e transformations canonical after having relaxed the parity conditions in the Yang-Mills case.

\section{Conclusions} \label{sec:conclusions}
In this paper, we have studied Yang-Mills theory  
with a particular focus on the fall-off and parity 
conditions that are needed in order give it a 
Hamiltonian formulation. Amongst the required 
structures is foremost the symplectic structure itself, 
a finite and functionally-differentiable Hamiltonian, 
and a canonical action of the Poincar\'e group.
Our aim was to find out to what extent these 
requirements allow for non-trivial groups of 
asymptotic symmetries and globally charges states,
fully analogous in spirit and technique to the 
corresponding investigations by Henneaux and Troessaert
for electrodynamics~\cite{Henneaux-ED}, 
gravity~\cite{Henneaux-GR}, and in the combination 
of the two~\cite{Henneaux-ED-GR}.

The fall-off conditions can be unequivocally determined
from a power-law ansatz if one requires that the usual
action of the Poincar\'e transformations leaves them
invariant. The discussion on the parity
conditions is more involved, as was expected from 
the experience with the electromagnetic case. 

We started by showing that strict parity conditions 
can be employed which allow the theory to meet all 
the required Hamiltonian requirements, though 
they turned out to not allow for improper gauge
transformations and non-zero global charges. 

We certainly did expect some additional constraints 
on the range of such conditions, over and above
those already known from the electrodynamic case. 
After all, there are additional terms from the 
non-vanishing commutators in the covariant derivatives which 
one needs to take care of. But we did not quite 
expect these constraints to be as restricting 
as they finally turned to be.  

In a second step we investigated into the 
possibility to regain non-trivial asymptotic 
symmetries and colour charges by carefully 
relaxing the parity conditions. We found that 
it is possible to relax the parity conditions
so that they are still preserved under Poincar\'e transformations, that the symplectic form 
is still finite, and that non-trivial improper 
gauge transformations exist. But this possibility 
had two independent drawbacks: 
First, the Poincar\'e transformations ceased to 
be canonical. We originally expected to be able 
to fix this issue in a manner similar to that 
employed in the electromagnetic case 
in~\cite{Henneaux-ED}, but this turned out 
not to work. Second, the relaxed parity 
conditions allowing non-zero colour charge 
fail to ensure the existence of a symplectic 
form. 

Let us clearly state that we do not pretend to 
have proven the impossibility of non-trivial 
asymptotic symmetries and non-vanishing 
global charges in an entirely rigorous sense,
taking full account of functional-analytic 
formulations of infinite-dimensional symplectic 
manifolds. However, the constraints we encountered 
are not of the kind that one can expect to 
simply disappear through proper identifications 
of function spaces. We believe that the obstructions 
we encountered point towards a deeper structural 
property of non-abelian Yang-Mills theory that has 
hitherto not been taken properly into consideration,
despite the fact that similar concerns 
were already raised  
several years ago in~\cite[Sec.~5]{Christodoulou.Murchadha:1981}
based on a careful asymptotic analysis of the 
field equations. Given that this view is 
correct, it is tempting to speculate that further 
clarification of that structure might tell us 
something relevant in connection with the problem 
of confinement.  After all, the general idea that 
confinement might be related to structures
already seen at a purely classical level is not new;
see, e.g., \cite{Feynman:1981}.  
  
An important further step would be to reconcile 
the Hamiltonian treatment at spacelike infinity
with the already existing study at null 
infinity~\cite{Strominger-YM,Barnich-YM}. Here, too, 
a confirmation of the obstructions we have seen 
would highlight a clear difference between
non-abelian Yang-Mills theory on one hand, and
electrodynamics and gravity on the other.
In particular, it would be of interest to learn 
whether such a reconciliation is possible only
at the price of allowing certain symmetries to
act non canonically.

\acknowledgments
% \section*{Acknowledgements}
% \noindent
We thank Marc Henneaux and 
C{\'e}dric Troessaert for conversations 
on the topic of this paper. 
Support by the DFG Research Training Group 1620 ``Models of Gravity'' is gratefully acknowledged. Finally we thank the anonymous referee for pointing out an omission in our previous version of 
formula \eqref{log-div1}, which led us 
to include the explicit calculation in 
appendix\,\ref{app:symplectic-form}.

\appendix
\section{The logarithmically-divergent contribution to the symplectic form} \label{app:symplectic-form}

In this appendix, we present a step-by-step computation of the logarithmically-divergent contribution to the symplectic form, which arises once we relax the  parity conditions to match~(\ref{parity-conditions-loosen1}) and (\ref{parity-conditions-loosen2}), as discussed in subsection~\ref{subsec:loosen-parity}.
In short, we will evaluate
\begin{equation} \label{log-div-symplectic}
 \begin{split}
  \oint_{S^2} d^2 \barr{x} \;
  \sprod{\extder \barr{\pi}^a \wedge}{ \extder \barr{A}_a} = 
  \oint_{S^2} d^2 \barr{x} \;
  \bigg[
  &\sprod{\extder \Big( \barr{\mathcal{U}}^{-1} \barr{\pi}^r_{\text{odd}} \, \barr{\mathcal{U}} \Big) \wedge}
  {\extder \Big( \barr{\mathcal{U}}^{-1} \barr{A}_r^{\text{even}} \, \barr{\mathcal{U}} \Big)}
  + \\
  +&\sprod{\extder \Big( \barr{\mathcal{U}}^{-1} \barr{\pi}^{\bar{a}}_{\text{even}} \, \barr{\mathcal{U}} \Big) \wedge}
  {\extder \Big( \barr{\mathcal{U}}^{-1} \barr{A}_{\bar{a}}^{\text{odd}} \barr{\mathcal{U}} \Big)}
  + \\
  +&\sprod{\extder \Big( \barr{\mathcal{U}}^{-1} \barr{\pi}^{\bar{a}}_{\text{even}} \, \barr{\mathcal{U}} \Big) \wedge}
  { \extder \Big( \barr{\mathcal{U}}^{-1} \partial_{\bar a}\, \barr{\mathcal{U}} \Big)}
  \bigg] \,.
 \end{split}
\end{equation}
Let us call $\barr{\Omega}_1$, $\barr{\Omega}_2$, and $\barr{\Omega}_3$ the contributions of the first, the second, and the third summand of the above expression, respectively.
In the following, we compute these three contributions separately.

\subsection{Preliminaries}
In order to make the ensuing computation of the three contributions easier to follow, let us evaluate in advance a few useful quantities.
To begin with, we note that most of the contributions in~(\ref{log-div-symplectic}) are of the form
\begin{equation} \label{div-gen1}
 \begin{aligned}
  \extder \Big( \barr{\mathcal{U}}^{-1} \eff \, \barr{\mathcal{U}} \, \Big) = &
  \, \extder \big( \barr{\mathcal{U}}^{-1} \big) \eff \, \barr{\mathcal{U}}+
  \barr{\mathcal{U}}^{-1} \extder \eff \, \barr{\mathcal{U}}+
  \barr{\mathcal{U}}^{-1} \eff \, \extder \barr{\mathcal{U}}= \\
  = & \, \barr{\mathcal{U}}^{-1}
  \Big( \extder \eff +
  \eff \, \extder \barr{\mathcal{U}} \, \barr{\mathcal{U}}^{-1}
  -\extder \barr{\mathcal{U}} \, \barr{\mathcal{U}}^{-1} \eff
  \Big) \,\barr{\mathcal{U}} = \\
  = & \, \barr{\mathcal{U}}^{-1}
  \Big( \extder \eff +
  \extprod{\eff}{\big(\extder \barr{\mathcal{U}} \, \barr{\mathcal{U}}^{-1} \big) }
  \Big)  \,\barr{\mathcal{U}} \,,
 \end{aligned}
\end{equation}
where $\eff$ needs to be replaced by one of definite-parity parts appearing in the canonical fields.
In the above expression, we have made use of the identity
$\extder \big( \barr{\mathcal{U}}^{-1} \big) =
- \barr{\mathcal{U}}^{-1} \extder \barr{\mathcal{U}} \; \barr{\mathcal{U}}^{-1}$, in order to obtain the expression on the second line.
Moreover, let us also compute
\begin{equation} \label{div-gen2}
 \begin{aligned}
  \extder \Big( \barr{\mathcal{U}}^{-1} \partial_{\bar a}\, \barr{\mathcal{U}} \Big) = &
  \, \extder \big( \barr{\mathcal{U}}^{-1} \big) \partial_{\bar a} \, \barr{\mathcal{U}}+
  \barr{\mathcal{U}}^{-1} \partial_{\bar a} \big( \extder \barr{\mathcal{U}} \big)= \\
  = & \, \barr{\mathcal{U}}^{-1} \Big[
  -\extder \barr{\mathcal{U}} \; \barr{\mathcal{U}}^{-1} \partial_{\bar a}\, \barr{\mathcal{U}} \; \barr{\mathcal{U}}^{-1} +
  \partial_{\bar a} \big( \extder \barr{\mathcal{U}} \big) \, \barr{\mathcal{U}}^{-1}
  \Big] \,\barr{\mathcal{U}} = \\
  = & \, \barr{\mathcal{U}}^{-1} \Big[
  \extder \barr{\mathcal{U}} \; \partial_{\bar a} \big( \barr{\mathcal{U}}^{-1} \big) +
  \partial_{\bar a} \big( \extder \barr{\mathcal{U}} \big) \, \barr{\mathcal{U}}^{-1}
  \Big] \,\barr{\mathcal{U}} = \\
  = & \, \barr{\mathcal{U}}^{-1}
  \partial_{\bar a} \Big(
  \extder \barr{\mathcal{U}} \;  \barr{\mathcal{U}}^{-1}
  \Big) \,\barr{\mathcal{U}} \,,
 \end{aligned}
\end{equation}
where we have made use of the further identity
$\partial_{\bar a} \big( \barr{\mathcal{U}}^{-1} \big) =
- \barr{\mathcal{U}}^{-1} \partial_{\bar a} \, \barr{\mathcal{U}} \; \barr{\mathcal{U}}^{-1}$,
in order to obtain the expression on the third line.

%%%%%%%% Nico's method %%%%%%%
Finally, let us evaluate
$\sprod{ \big( \extprod{U}{\omega} \big) \wedge }
{\big( \extprod{V}{\omega} \big)} $,
where $U$ and $V$ are $\su(N)$-valued functions 
and $\omega$ is a $\su(N)$-valued one-form on phase 
space to which the exterior product refers.
From our definition \eqref{eq:CompModifiedKillingProduct} 
of the inner product, we get
\begin{equation}
 \sprod{ \big( \extprod{U}{\omega} \big) \wedge }{\big( \extprod{V}{\omega} \big)} =
 - \tr \Big( 
 \extprod{U}{\omega} \wedge \extprod{V}{\omega}
 \Big) =
 -\tr \Big( 
 \big[ U , \omega \big] \wedge 
 \big[ V , \omega \big]
 \Big) \,,
\end{equation}
where next to the exterior product of $\su(N)$-valued 
one-forms matrix multiplication in $\su(N)$ is also understood. In the following we shall also temporarily 
drop the wedge-product symbol. We only need to remember 
to insert an extra minus sign every time we invert 
the order of the two $\omega$.
Expanding the commutators and the composition, we get
\begin{equation}
 \sprod{ \big( \extprod{U}{\omega} \big) \wedge }{\big( \extprod{V}{\omega} \big)} =
 -\tr \Big( 
 U \omega V \omega +
 \omega U \omega V -
 U \omega \omega V -
 \omega UV\omega
 \Big) \,.
\end{equation}
Using the cyclicity of the trace and taking
into account the minus sign whenever the order of the two one-forms 
$\omega$ changes, we immediately see that the first two terms cancel. 
Therefore, applying the same rules, we get
\begin{equation}
 \sprod{ \big( \extprod{U}{\omega} \big) \wedge }{\big( \extprod{V}{\omega} \big)} =
 - \tr \Big( 
 - U \omega \omega V
 - \omega U V \omega
 \Big) =
 - \tr \Big( 
 - \omega \omega V U
 + \omega \omega U V
 \Big)
 \,.
\end{equation}
This can be factorised in the form 
\begin{equation}
 \sprod{ \big( \extprod{U}{\omega} \big) \wedge }{\big( \extprod{V}{\omega} \big)} =
 - \tr \Big( 
 \omega \omega \big[ U , V \big]
 \Big) =
 - \tr \left( 
 \frac{1}{2} \big[ \omega , \omega \big]
 \, \big[ U , V \big]
 \right)
 \,,
\end{equation}
where we have replaced the product $\omega\omega$ with the commutator divided by two using the antisymmetry of the exterior product.\footnote{Note that the commutator of the two $\omega$ does not identically vanish because it is combined with the (antisymmetric) exterior product.}
Finally, recalling our definition \eqref{eq:CompModifiedKillingProduct} of the inner 
product and again displaying the exterior product, 
we arrive at the desired identity
\begin{equation} \label{double-product-identity}
 \sprod{ \big( \extprod{U}{\omega} \big) \wedge }{\big( \extprod{V}{\omega} \big)} =
 \frac{1}{2}
 \sprod{
 \big( \extprod{\omega \wedge}{\omega} \big)
 }{
 \big( \extprod{U}{V} \big)
 } \,.
\end{equation}
We are now ready to present the actual computation of the three terms $\barr{\Omega}_1$, $\barr{\Omega}_2$, and $\barr{\Omega}_3$, whose sum gives the divergent contribution~(\ref{log-div-symplectic}) to the symplectic form.

\subsection{Computation of the divergent contribution}
First, let us compute $\barr{\Omega}_1$, the first line of the right-hand side of~(\ref{log-div-symplectic}).
Using~(\ref{div-gen1}), we get
\begin{equation}
 \begin{aligned}
  \barr{\Omega}_1 \eqdef
  & \oint_{S^2} \!\! d^2 \barr{x} \;
  \sprod{\extder \Big( \barr{\mathcal{U}}^{-1} \barr{\pi}^r_{\text{odd}} \, \barr{\mathcal{U}} \Big) \wedge}
  {\extder \Big( \barr{\mathcal{U}}^{-1} \barr{A}_r^{\text{even}} \barr{\mathcal{U}} \Big)} = \\
  =& \oint_{S^2} \!\! d^2 \barr{x} \;
  \sprod
  {\barr{\mathcal{U}}^{-1}
  \Big[ \extder \barr{\pi}^r_{\text{odd}} +
  \extprod{\barr{\pi}^r_{\text{odd}}}{\Big(\extder \barr{\mathcal{U}} \, \barr{\mathcal{U}}^{-1} \Big) }
  \Big]  \,\barr{\mathcal{U}} \, \wedge}
  {\barr{\mathcal{U}}^{-1}
  \Big[ \extder \barr{A}_r^{\text{even}} +
  \extprod{\barr{A}_r^{\text{even}}}{\Big(\extder \barr{\mathcal{U}} \, \barr{\mathcal{U}}^{-1} \Big) }
  \Big]  \,\barr{\mathcal{U}}} = \\
  =& \oint_{S^2} \!\! d^2 \barr{x} \;
  \sprod
  {\Big[ \extder \barr{\pi}^r_{\text{odd}} +
  \extprod{\barr{\pi}^r_{\text{odd}}}{\Big(\extder \barr{\mathcal{U}} \, \barr{\mathcal{U}}^{-1} \Big) }
  \Big]  \wedge}
  {\Big[ \extder \barr{A}_r^{\text{even}} +
  \extprod{\barr{A}_r^{\text{even}}}{\Big(\extder \barr{\mathcal{U}} \, \barr{\mathcal{U}}^{-1} \Big) }
  \Big]  } \,,
 \end{aligned}
\end{equation}
where, on the last step, we have simplified $\barr{\mathcal{U}}$ with $\barr{\mathcal{U}}^{-1}$ using the cyclicity of the trace, which appears in the definition of the Killing inner product.
At this point, we can expand the product of the two terms in square brackets.
The term
$\sprod{\extder \barr{\pi}^r_{\text{odd}} \wedge}{\extder \barr{A}_r^{\text{even}}}$
vanishes upon integration because it is an odd function on the sphere.
Using the symmetries of the triple product and being careful in putting an extra minus sign every time we change the order of the forms in the exterior product, we can rearrange the terms as
\begin{equation} \label{omega1-partial}
 \begin{aligned}
  \barr{\Omega}_1
  = \oint_{S^2} d^2 \barr{x} \;
  &\Bigg\{
  \sprod{\Big(\extder \barr{\mathcal{U}} \, \barr{\mathcal{U}}^{-1} \Big) \wedge}
  { \Big(
  \extprod{ \barr{A}_r^{\text{even}} }{\extder \barr{\pi}^r_{\text{odd}}}+
  \extprod{\extder \barr{A}_r^{\text{even}} }{ \barr{\pi}^r_{\text{odd}}}
  \Big) }
  + \\
  &+\sprod
  {\Big[
  \extprod{\barr{\pi}^r_{\text{odd}}}{\Big(\extder \barr{\mathcal{U}} \, \barr{\mathcal{U}}^{-1} \Big) }
  \Big]  \wedge}
  {\Big[
  \extprod{\barr{A}_r^{\text{even}}}{\Big(\extder \barr{\mathcal{U}} \, \barr{\mathcal{U}}^{-1} \Big) }
  \Big]}
  \Bigg\} \,.
 \end{aligned}
\end{equation}
The second factor in the first summand can be rewritten as
$\extder \big(
\extprod{ \barr{A}_r^{\text{even}} }{ \barr{\pi}^r_{\text{odd}}}
\big)$, simply using the Leibniz rule.
Moreover, the second summand can be rewritten using the identity~(\ref{double-product-identity}).
Hence, we arrive at the expression
\begin{equation} \label{div-line1}
 \begin{aligned}
  \barr{\Omega}_1
  =\oint_{S^2} d^2 \barr{x} \;
  &\bigg\{
  \sprod{\Big(
  \extder \barr{\mathcal{U}} \;  \barr{\mathcal{U}}^{-1}
  \Big) \wedge }
  {\extder \Big(
  \extprod{\barr{A}_r^{\text{even}}}{\barr{\pi}^r_{\text{odd}}}
  \Big)}+ \\
  &-\sprod{
  \frac{1}{2} \Big[\extprod{ \Big(
  \extder \barr{\mathcal{U}} \, \barr{\mathcal{U}}^{-1} \Big) \wedge }{ \Big(\extder \barr{\mathcal{U}} \, \barr{\mathcal{U}}^{-1} \Big)} \Big] 
   }
  { \Big( \extprod{\barr{A}_r^{\text{even}}}{\barr{\pi}^r_{\text{odd}}} \Big) }
  \bigg\} \,.
 \end{aligned}
\end{equation}

Second, let us note that the second line of~(\ref{log-div-symplectic}) is analogous to the first line, so that we can get the value of $\barr{\Omega}_2$ with a computation almost identical to the one for $\barr{\Omega}_1$, obtaining
\begin{equation} \label{div-line2}
 \begin{aligned}
  \barr{\Omega}_2
  = \oint_{S^2} d^2 \barr{x} \;
  &\bigg\{ 
  \sprod{\Big(
  \extder \barr{\mathcal{U}} \;  \barr{\mathcal{U}}^{-1}
  \Big) \wedge }
  {\extder \Big(
  \extprod{\barr{A}_{\bar a}^{\text{odd}}}{\barr{\pi}^{\bar a}_{\text{even}}}
  \Big)} +\\
  &-\sprod{
  \frac{1}{2} \Big[\extprod{ \Big(
  \extder \barr{\mathcal{U}} \, \barr{\mathcal{U}}^{-1} \Big) \wedge }{\Big(\extder \barr{\mathcal{U}} \, \barr{\mathcal{U}}^{-1} \Big)} \Big] 
   }
  { \Big( \extprod{\barr{A}_{\bar a}^{\text{odd}}}{\barr{\pi}^{\bar a}_{\text{even}}} \Big) }
  \bigg\} \,.
 \end{aligned}
\end{equation}

Third, let us compute the last contribution $\barr{\Omega}_3$.
Using~(\ref{div-gen1}) and~(\ref{div-gen2}), we get
\begin{equation}
 \begin{aligned}
  \barr{\Omega}_3 \eqdef
  & \oint_{S^2} d^2 \barr{x} \;
  \sprod{\extder \Big( \barr{\mathcal{U}}^{-1} \barr{\pi}^{\bar{a}}_{\text{even}} \, \barr{\mathcal{U}} \Big) \wedge}
  { \extder \Big( \barr{\mathcal{U}}^{-1} \partial_{\bar a}\, \barr{\mathcal{U}} \Big)}
  = \\
  = & \oint_{S^2} d^2 \barr{x} \; %Line 2
  \sprod
  {\barr{\mathcal{U}}^{-1}
  \Big[ \extder \barr{\pi}^{\bar a}_{\text{even}} +
  \extprod{\barr{\pi}^{\bar a}_{\text{even}}}{\Big(\extder \barr{\mathcal{U}} \, \barr{\mathcal{U}}^{-1} \Big) }
  \Big]  \,\barr{\mathcal{U}} \;\wedge}
  {\barr{\mathcal{U}}^{-1}
  \partial_{\bar a} \Big(
  \extder \barr{\mathcal{U}} \;  \barr{\mathcal{U}}^{-1}
  \Big) \,\barr{\mathcal{U}} } \,.
 \end{aligned}
\end{equation}
Once again, we can simplify $\barr{\mathcal{U}}$ and $\barr{\mathcal{U}}^{-1}$ using the cyclicity of the trace employed in the definition of the Killing  inner product.
Expanding afterwards the expression, we get
\begin{equation}
 \begin{aligned}
  \barr{\Omega}_3
  = & \oint_{S^2} d^2 \barr{x} \;
  \bigg\{
  \sprod{\extder \barr{\pi}^{\bar a}_{\text{even}} \, \wedge}
  {\partial_{\bar a} \Big(
  \extder \barr{\mathcal{U}} \;  \barr{\mathcal{U}}^{-1}
  \Big)} +
  \sprod{
  \Big[\extprod{\barr{\pi}^{\bar a}_{\text{even}}}{\Big(\extder \barr{\mathcal{U}} \, \barr{\mathcal{U}}^{-1} \Big)} \Big] \wedge
  }
  {\partial_{\bar a} \Big(
  \extder \barr{\mathcal{U}} \;  \barr{\mathcal{U}}^{-1}
  \Big)}
  \bigg\} = \\
  = & \oint_{S^2} d^2 \barr{x} \;
  \bigg\{
  - \sprod{\extder \Big( \partial_{\bar a} \barr{\pi}^{\bar a}_{\text{even}} \Big) \, \wedge}
  {\Big(
  \extder \barr{\mathcal{U}} \;  \barr{\mathcal{U}}^{-1}
  \Big)} +
  \sprod{
  \Big[\extprod{\partial_{\bar a} \Big(
  \extder \barr{\mathcal{U}} \; \barr{\mathcal{U}}^{-1} \Big) \wedge}{\Big(\extder \barr{\mathcal{U}} \, \barr{\mathcal{U}}^{-1} \Big)} \Big] 
  }
  {\barr{\pi}^{\bar a}_{\text{even}}}
  \bigg\} \,,
 \end{aligned}
\end{equation}
where we have integrated by part the first summand.
Moreover, in the second summand, we have used the symmetries of the triple product and inserted an extra minus sign due to the ordering of the forms in the exterior product.
The above expression can be easily rewritten as
\begin{equation} \label{div-line3}
 \begin{aligned}
  \barr{\Omega}_3
  = & \oint_{S^2} d^2 \barr{x} \;
  \bigg\{
  \sprod{\Big(
  \extder \barr{\mathcal{U}} \;  \barr{\mathcal{U}}^{-1}
  \Big) \wedge}
  {\extder \Big( \partial_{\bar a} \barr{\pi}^{\bar a}_{\text{even}} \Big)}+
  \sprod{
  \frac{1}{2} \partial_{\bar a} \Big[\extprod{ \Big(
  \extder \barr{\mathcal{U}} \; \barr{\mathcal{U}}^{-1} \Big) \wedge}{\Big(\extder \barr{\mathcal{U}} \, \barr{\mathcal{U}}^{-1} \Big)} \Big] 
  }
  {\barr{\pi}^{\bar a}_{\text{even}}}
  \bigg\} = \\
  = & \oint_{S^2} d^2 \barr{x} \;
  \bigg\{
  \sprod{\Big(
  \extder \barr{\mathcal{U}} \;  \barr{\mathcal{U}}^{-1}
  \Big) \wedge}
  {\extder \Big( \partial_{\bar a} \barr{\pi}^{\bar a}_{\text{even}} \Big)}-
  \sprod{
  \frac{1}{2} \Big[\extprod{ \Big(
  \extder \barr{\mathcal{U}} \; \barr{\mathcal{U}}^{-1} \Big) \wedge}{\Big(\extder \barr{\mathcal{U}} \, \barr{\mathcal{U}}^{-1} \Big)} \Big] 
  }
  { \partial_{\bar a} \barr{\pi}^{\bar a}_{\text{even}} }
  \bigg\} \,,
 \end{aligned}
\end{equation}
where we have integrated by part the second summand.

Finally, we find the logarithmically-divergent contribution to the symplectic form by summing the three contributions $\barr{\Omega}_1$, $\barr{\Omega}_2$, and $\barr{\Omega}_3$, given by the expressions~(\ref{div-line1}), (\ref{div-line2}), and~(\ref{div-line3}), respectively.
The result coincides exactly with the expression~(\ref{log-div1}) presented in subsection~\ref{subsec:loosen-parity}.

\section{Details about the computations of section~\ref{subsec:attempt-Poincare-canonical}} \label{app:details-computations}
In this appendix, we provide a more detailed discussion about the attempts to make the Poincar\'e transformations canonical after having relaxed the parity conditions in the Yang-Mills case.
In particular, we extend the information of subsection~\ref{subsubsec:ansatz}.
There, some assumptions were made in the behaviour of the fields under Poincar\'e transformations and in the ansatz~(\ref{ansatz-omega}) for the boundary term of the symplectic form.
In the following, we comment on the fact that these assumptions are actually not so restrictive.

\subsection{Poincar\'e transformations of the fields}
We remind that, in section~\ref{subsubsec:ansatz}, we introduced a one form $\phi_a$ and the conjugated momenta $\Pi^a$.
These new canonical fields were required to satisfy the fall-off conditions~\ref{fall-off-phi} and the further constraint $\Pi^a \approx 0$.
At this point, we have to specify how the fields transform under the Poincar\'e transformations and, in particular, the Lorentz boost.
In order to do so, let us make a few assumptions.

First, we wish that, ultimately, the Poincar\'e transformations would be generated by the Poisson brackets with a function $P$ on phase space, as it is in the case of general relativity and electrodynamics.
So, let us write the candidate for the generator of the boost as
\begin{equation} \label{boost-generator-new0}
 P' [\xi^\perp] \eqdef
 \int d^3 x \, \xi^\perp \left[
 \frac{\sprod{\pi^a}{\pi_a}}{2\sqrt{g}} + \frac{\sqrt{g}}{4} \sprod{F_{ab}}{F^{ab}}
 +\mathscr{P}'_{(1)}
 %- \sprod{\Psi}{\mathscr{G}} - \sprod{A_a}{\nabla^a \pi_\Psi}
 \right]
 + (\text{boundary}) \,,
\end{equation}
where the first two summands in the square brackets are responsible for the usual transformations~(\ref{poincare-transformations1}) and~(\ref{poincare-transformations2}), while $\mathscr{P}'_{(1)}$ takes into account the transformation of $\phi_a$ and $\Pi^a$, as well as some possible new contributions to the transformation of $A_a$ and $\pi^a$.
For now, we ignore any issue concerning the existence of a boundary term which makes the generator above well defined.
We pretend that it exists, in order to allow the following formal manipulations, and check at the end whether or not this is consistent.
It is the goal of this appendix to show that such boundary term does \emph{not} actually exist.
Note that, due to the presence of an unspecified boundary term in the expression above, $\mathscr{P}'_{(1)}$ is defined up to a total derivative.
We will implicitly make use of this fact in some of the following equalities.

Secondly, the attempt done in section~\ref{subsubsec:ansatz-Psi} failed because it was not possible to compensate for the term containing
$\sprod{\extder \barr{A}_{\bar m} \,\wedge}{ \extder( \extprod{ \barr{A}^{\bar m} }{\barr{A}_r} ) }$
in $\liephase_{X} \Omega$.
Indeed, there was no field transforming (asymptotically) as (the asymptotic part of) $A_a$ without any derivative.
Therefore, as a further assumption, we ask that $\phi_a$ transforms exactly as $\delta_{\xi^\perp} \phi_a = \xi^\perp A_a$, thus finding
\begin{equation} \label{boost-generator-new1}
  \mathscr{P}'_{(1)} = \sprod{A_a}{\Pi^a} + \mathscr{P}'_{(2)} \,,
\end{equation}
where $\mathscr{P}'_{(2)}$ does \emph{not} depend on $\Pi^a$.

Thirdly, we ask that the transformations of $A_a$ and $\pi^a$ differ from the ones in~(\ref{poincare-transformations1}) and~(\ref{poincare-transformations2}) by, at most, gauge transformations and constraints.
Since $\mathscr{P}'_{(2)}$ cannot depend on the constraints $\Pi^a \approx 0$, this implies that $\mathscr{P}'_{(2)} = \sprod{F}{\mathscr{G}}$, for some function $F$ of the canonical fields (except for $\Pi^a$) and their derivatives.
Note that, since $\mathscr{G}$ is a weight-one scalar density, $F$ needs to be a scalar in order for the integral in~(\ref{boost-generator-new0}) to make sense.

Finally, we require that the transformations of $A_a$ and $\pi^a$ are exactly the ones in~(\ref{poincare-transformations1}) and~(\ref{poincare-transformations2}), when the new fields $\phi_a$ and $\Pi^a$ are set to zero.
Therefore, we can write, up to boundary terms, $F = \mathscr{D}^a \phi_a$, where the operator $\mathscr{D}^a$ is built using the fields $A_a$, $\pi^a$, and $\phi_a$, as well as an arbitrary number of derivatives and $\su(N)$ commutators.
At the lowest order in the derivatives and in the fields, we find
\begin{equation}
 \mathscr{D}^a \phi_a = c_0 \nabla^a \phi_a + c_1 \extprod{A^a}{\phi_a} + c_2 \extprod{\pi^a}{\phi_a} \,,
\end{equation}
where $c_0, c_1, c_2 \in \real$ are three free parameters.
After noting that $c_0$ can be set to $1$ by redefining $\phi_a$, we find exactly the transformations~(\ref{boost-YM-phi-in})--(\ref{boost-YM-phi-end}), that were assumed in section~\ref{subsubsec:ansatz}.

\subsection{The boundary term of the symplectic form}

Before we can verify whether or not the Poincar\'e transformations are canonical, we need to specify how the symplectic form is affected by the introduction of the new fields $\phi_a$ and $\Pi^a$.
We assume that the contribution in the bulk is of the usual form $\sprod{\extder \Pi^a \, \wedge}{\extder \phi_a}$.
Therefore, the symplectic form in the bulk $\Omega'$ is given by~(\ref{Omega-bulk-phi}).
To this, we add a boundary term $\omega'$ built using the asymptotic part of the fields.
There is potentially an infinite number of possibilities when one writes contributions to $\omega'$.
However, a few things need to be taken into consideration.

First, we want to achieve $\liephase_{X'} (\Omega' + \omega') = 0 $.
Now, $\liephase_{X'} \Omega'$ contains non-zero boundary contributions as shown in~(\ref{lie-symplectic-bulk}).
In order for $\liephase_{X'} (\Omega' + \omega')$ to be actually zero, we need that terms in~(\ref{lie-symplectic-bulk}) are compensated by some terms in $\liephase_{X'} \omega'$.
The ansatz~(\ref{ansatz-omega}) is designed exactly in this spirit.
In particular, the terms with coefficients $a_0$ and $a_1$ should compensate those parts of~(\ref{lie-symplectic-bulk}) containing derivatives of $\barr{A}$ and those containing $\barr{\pi}^r$, whereas the terms with coefficients $a_2,\dots,a_6$ should tackle the part in~(\ref{lie-symplectic-bulk}) containing
$\sprod{\extder \barr{A}_{\bar m} \,\wedge}{ \extder( \extprod{ \barr{A}^{\bar m} }{\barr{A}_r} ) }$.

Secondly, introducing contributions to $\omega'$ built using the momenta $\barr{\pi}^a$ does not help, since these would introduce terms with at least two derivatives of $\barr{A}_a$ in $\liephase_{X'} \omega'$, due to their asymptotic transformations under boost --- see~(\ref{poincare-asymptotic3}) and~(\ref{poincare-asymptotic4}) --- while $\liephase_{X'} \Omega'$ only contains terms with at most one derivative.

Thirdly, having terms in $\omega'$ containing a great number of fields and of their commutators would introduce a big complication in the problem.
Furthermore, it would be difficult to justify such terms when comparing the theory at spatial and at null infinity.

In conclusion, we consider the ansatz~(\ref{ansatz-omega}) for the boundary term of the symplectic form for the aforementioned reasons.
Although it is not the most general ansatz, it is general enough to show that the Yang-Mills case is substantially different from electrodynamics and general relativity.

\subsection{The Poincar\'e transformations are not canonical}
We finally show that no value of the free parameters $a_0,\dots,a_6$, $c_1$, and $c_2$ makes the Poincar\'e transformations canonical.
To begin with, the symplectic form $\Omega' + \omega'$ must be a closed two-form on phase space.
Since $\extder \Omega' = 0$, one need to impose that also $\extder \omega' = 0$.
One can easily check that this amount to consider the general ansatz~(\ref{ansatz-omega}) with the free parameters $a_0,\dots,a_6$ restricted by the two conditions
\begin{align} \label{conditions-param-1}
 a_3 + a_4 = 0
 && \text{and} &&
 a_2 + a_5 + a_6 = 0 \,.
\end{align}
The two conditions above imply that one can rewrite the boundary term $\omega'$ of the symplectic form as
\begin{equation} 
\label{ansatz-omega-close}
 \begin{aligned}
 \omega' &= \oint_{S^2} d^2 \barr{x} \; \sqrt{\barr{\gamma}} \,
 \Big[
  a_0 \, \sprod{\extder \big( \barr{\nabla}^{\bar m}  \barr{\phi}_{\bar m} \big) \wedge}{\extder \barr{A}_r } 
 +a_1 \, \sprod{\extder \barr{\phi}_{r} \wedge}{\extder \barr{A}_r } \\
 + & {\tilde a}_2 \, \sprod{ \extder \barr{A}_{\bar m} \, \wedge }{ \extder \big( \extprod{ \barr{A}^{\bar m} }{ \barr{\phi}_r } \big) }
 + {\tilde a}_3 \, \sprod{ \extder \barr{A}_{\bar m} \, \wedge }{ \extder \big( \extprod{ \barr{A}_r }{ \barr{\phi}^{\bar m} } \big) }
 + {\tilde a}_4 \, \sprod{ \extder \barr{A}_{r} \, \wedge }{ \extder \big( \extprod{ \barr{A}_{\bar m} }{ \barr{\phi}^{\bar m} } \big) }
 \Big] \,,
\end{aligned}
\end{equation}
where the three parameters ${\tilde a}_2$, ${\tilde a}_3$, and ${\tilde a}_4$ are related to $a_2, \dots, a_5$ by
\begin{align} \label{conditions-param-2}
 \tilde a_2 = a_3 \,, &&
 \tilde a_3 = -a_2 \,, \text{ and} &&
 \tilde a_4 = -a_5 \,.
\end{align}
Note that~(\ref{ansatz-omega-close}) is not only close but also exact.

It is now not difficult to show that the Poincar\'e transformations are not canonical for any value of the free parameters $a_0$, $a_1$, ${\tilde a}_2$, ${\tilde a}_3$, ${\tilde a}_4$, $c_1$, and $c_2$.
Indeed, the Poincar\'e transformations would be canonical if, and only if,
\begin{equation} \label{lie-Omega-omega}
 \liephase_{X'} \Omega' = - \liephase_{X'} \omega' \,.
\end{equation}
The left-hand side of the above expression was already computed in~(\ref{lie-symplectic-bulk}).
It contains a first summand with the term
$\sprod{\extder \barr{A}_{\bar m} \,\wedge}{ \extder \big( D^{\bar m} \barr{A}_r \big) }$,
which would appear also on the right-hand side of the above expression if we imposed
\begin{align} \label{conditions-param-3}
 a_0 = 1
 && \text{and} &&
 {\tilde a}_3 = {\tilde a}_2 +1 \,.
\end{align}
Moreover, the left-hand side of~(\ref{lie-Omega-omega}) contains a second summand with the term
$\sprod{ \extder \barr{\pi}^r \, \wedge}{ \extder \barr{\mathscr{D} \phi }}$.
This contribution would be compensated by a similar contribution on the right-hand side of~(\ref{lie-Omega-omega}) if we imposed the further conditions
\begin{align} \label{conditions-param-4}
 a_1 = 1 \,, &&
 c_1 = 0 \,, \text{ and} &&
 c_2 = 0 \,.
\end{align}
After restricting the free parameters to those satisfying~(\ref{conditions-param-3}) and~(\ref{conditions-param-4}), every term in the left-hand side of~(\ref{lie-Omega-omega}) appears also on the right-hand side.
However, the latter contains also other terms, which one has to set to zero with an appropriate choice of the remaining parameters, if this is actually possible.
In particular, the right-hand side still contains, among others, some contribution proportionate to $\extder \barr{\pi}^r$.
These would vanish, if we set
\begin{align} \label{conditions-param-5}
 {\tilde a}_2 = -1
 && \text{and} &&
 {\tilde a}_4 = 0 \,.
\end{align}
These conditions, together with the previous ones~(\ref{conditions-param-1}), (\ref{conditions-param-2}), (\ref{conditions-param-3}), and (\ref{conditions-param-4}), completely fix the values of the free parameters, so that
\begin{equation} \label{ansatz-omega-final}
 \omega' = \oint_{S^2} d^2 \barr{x} \; \sqrt{\barr{\gamma}} \,
 \Big[
 \sprod{\extder \big( 2 \barr{\phi}_{r} + \barr{\nabla}^{\bar m}  \barr{\phi}_{\bar m} \big) \wedge}{\extder \barr{A}_r } 
 - \sprod{ \extder \barr{A}_{\bar m} \, \wedge }{ \extder \big( \extprod{ \barr{A}^{\bar m} }{ \barr{\phi}_r } \big) }
 \Big]
\end{equation}
does not depend any more on any free parameter, nor do the Poincar\'e transformations (\ref{boost-YM-phi-in})--(\ref{boost-YM-phi-end}).
One can now easily verify by direct computation that $\liephase_{X'} (\Omega' + \omega') \ne 0$, i.e., the Poincar\'e transformations are not canonical, as we wanted to show.
In particular, this also shows that the boundary term in~(\ref{boost-generator-new0}) cannot exist.

\bibliography{biblio}{}
 \bibliographystyle{jhep-fullnames}
% \begin{thebibliography}{00}
% \end{thebibliography}
\end{document}